\documentclass{article}

\usepackage{arxiv}

\usepackage[utf8]{inputenc} 
\usepackage[T1]{fontenc}    
\usepackage{hyperref}       
\usepackage{url}            
\usepackage{booktabs}       
\usepackage{amsfonts}       
\usepackage{nicefrac}       
\usepackage{microtype}      
\usepackage{lipsum}
\usepackage{graphicx}
\graphicspath{ {./images/} }

\usepackage{cite}
\usepackage{amsmath,amssymb,amsfonts}
\usepackage{textcomp}
\usepackage{comment}
\usepackage{algorithm}
\usepackage{algorithmic}

\usepackage{ragged2e}

\title{Interplay Between Hierarchy and Centrality in Complex Networks}

\author{
  Stephany Rajeh*, Marinette Savonnet, Eric Leclercq, and Hocine Cherifi \\
  Laboratoire d'Informatique de Bourgogne\\
  University of Burgundy\\
  21000 Dijon, France\\
  \texttt{*stephany.rajeh@u-bourgogne.fr} \\

}

\begin{document}
\maketitle

\begin{abstract}
Hierarchy and centrality are two popular notions used to characterize the importance of entities in complex systems. Indeed, many complex systems exhibit a natural hierarchical structure, and centrality is a fundamental characteristic allowing to identify key constituents. Several measures based on various aspects of network topology have been proposed in order to quantify these concepts. While numerous studies have investigated whether centrality measures convey redundant information, how centrality and hierarchy measures are related is still an open issue. In this paper, we investigate the interaction between centrality and hierarchy using several correlation and similarity evaluation measures. A series of experiments is performed in order to evaluate the combinations of 6 centrality measures with 4 hierarchy measures across 28 diverse real-world networks with varying topological characteristics. Results show that network density and transitivity play a key role in shaping the pattern of relations between centrality and hierarchy measures.
\end{abstract}


\section{Introduction}
\label{sec:introduction}
Networks offer a powerful representation of complex systems in which nodes represent the elementary units of the system and links represent their interactions. They are well adapted to investigate the relationships between structure and dynamics in such systems from the macroscopic to the microscopic level. At the microscopic level, the various roles of the nodes are induced by their specific pattern of connectivity. These interactions among nodes can cause them to exhibit varying levels of complexity. In turn, arising levels of complexity hinder the process of identifying the most influential nodes in a complex networks \cite{baker2013complexity}. Studying these interactions and quantifying the importance of the nodes has attracted a great number of researchers. Indeed, identifying key nodes is of prime interest in a wide range of strategic applications. Such applications span on fundamental fields such as prevention of epidemic spreading and vaccination strategies, understanding the origins of diseases, maximizing information diffusion of marketing campaigns, ensuring robust power grids, identification of climate change and ocean currents, finding key heads in terrorist networks, and many more \cite{survey2}. Although, characterizing  the importance  of the nodes in a network can be apprehended from various perspectives, the vast majority of works concern centrality and hierarchy.

Centrality is one of the main research topics in network science literature. Centrality measures quantify the ability of a node to influence the other nodes of the network based on the topological properties that are driving the network dynamics \cite{ghanem2018centrality}. Early work has identified the key aspects of centrality. First, the number of local connections of a node and its ability to spread locally the information. Second, the position of a node in the network and its importance in the global exchange of information. A multitude of centrality measures have been proposed to date \cite{survey1}. They can be classified into different groups. Indeed, they can be seen as global/path-based or local/neighborhood-based \cite{das2014local, lu2016vital}. They can be based on iterative-refinement process \cite{lu2016vital}. They can even be multidimensional by combining different measures together \cite{multidimensional} or by incorporating the influence of the community structure \cite{ghalmane2019centrality,ghalmane2019centrality1,cherifi2019community}.

Hierarchy is one of the foundations to understand complex networks, as they often take the form of hierarchical structure \cite{e1}. Indeed, hierarchy is ubiquitous throughout numerous natural or man-made complex networks. It can be observed in animal complex systems such as leader-follower network of pigeon flocks, biological complex networks such as neural networks, transportation networks such as road networks, within organizations and between organizations, in social networks, and even the way we speak \cite{zafeiris2017we, clark2013handbook}. One may ask why does hierarchy exist after all? Herbert Simon \cite{e1} tried to answer by his famous "watchmaker parable". It is having structure within systems such that each subsystem depends on another, and it results in more efficiency and resiliency than a random structure.

Several hierarchical measures based on different definitions of hierarchy have been developed \cite{e9, e8, e13, e14, e15, garcia2017ranking, wang2012truss}. Generally, they are used to quantify the hierarchical level of the whole network. Here,  the focus is on the hierarchical level of individual nodes. One can distinguish two type of hierarchy measures in this case: nested and flow hierarchy. In nested hierarchy higher level elements are contained in lower level elements while in flow hierarchy nodes are arranged in different levels such that influential nodes are at a higher level and they are connected to the nodes they influence. Main nested hierarchy measures ($k$-core and $k$-truss)  are based on hierarchical decomposition of nodes \cite{malliaros2020core, garcia2017ranking, wang2012truss}. To our knowledge, Local Reaching Centrality" (LRC)  is the only  flow hierarchy measure used to quantify node hierarchy \cite{e9}.

Although the notions of hierarchy and centrality are quite different from a sociological point of view \cite{Newman2010}, the distinction seems more subtle when it comes to the associated measures. Indeed, both try to quantify the "importance" of a node based on topological information. One can consider that the most influential node is the one with the highest number of connections, or that connects two communities with each other, or the one allowing to reach all neighbors the fastest. This may be the case in some networks, but high centrality nodes aren't necessarily the most strategically located in terms of hierarchy \cite{kitsak2010identification}. In theory, these measures should capture different roles in the network.
There have been some works on centrality measures incorporating some hierarchical aspects, such as social centrality, control centrality, and improved betweenness centrality \cite{saxena2018social, liu2012control, li2017hierarchical}. Additionally, there has been numerous studies in order to evaluate how various centrality measures are related \cite{oldham2019consistency}. However, up to now, to our knowledge, this work is the first attempt to investigate systematically how centrality and hierarchy measures are related and what is the impact of the network topology on this relationship.

To investigate this issue, a set of six measures spanning the three main types of centrality (neighborhood-based, path-based and iterative refinement-based) together with a set of four hierarchy measures representing three categories (nestedness based, flow based, and mixed based) are compared across a large set of real-world networks originating from various fields. Networks are chosen in order to span a wide range of basic topological property values such as density, transitivity and assortativity.
A systematic evaluation of the various combinations of centrality and hierarchy measures is performed using three correlation measures (Pearson, Spearman, Kendall Tau and two similarity measures (Jaccard, RBO) for every network in order to explore to which extent the various combinations convey different information. Variations of  these relations across networks are studied in order to clarify the relationship between hierarchy and centrality of nodes, and network topology.

The main contributions of this work is threefold:

\begin{enumerate}
    \item{A systematic investigation of how centrality measures are related to the main hierarchy measures in a number of real-world networks is performed;}
    
    \item{The interplay between hierarchy and centrality measures is related to the basic network topological properties;}
    
    \item{The most orthogonal hierarchy and centrality measures, whatever the network topological characteristics, are specified. }
\end{enumerate}

The rest of the paper is organized as follows. Section \ref{sec:Hierarchy} introduces the hierarchy measures under study. Section \ref{sec:Centrality} presents the centrality measures. The evaluation measures used to investigate the relationship between hierarchy and centrality are presented in section \ref{sec:evalmeasures}. Section \ref{sec:DataAndMethodology} provides an overview of the data sets used and describes the methods applied. Section \ref{sec:ExperimentalResults} is devoted to the results and section \ref{sec:discussion} develops the discussion. Finally, section \ref{sec:concl} concludes the article.

\section{Hierarchy}
\label{sec:Hierarchy}
The word \textit{hierarchy} originates from the Greek word  \textit{hierarch\=es}\footnote{Merriam-Webster Dictionary:\newline
\url{https://www.merriam-webster.com/dictionary/hierarchy}.}, which is comprised of \textit{hieros} meaning holy and \textit{archos} meaning ruler. From here, we can come to terms that this word justifies a kind of power, superiority, or importance.  The most basic definition is the following: \textit{\textbf {hierarchy} is a system in which people or things are arranged according to their importance}\footnote{Cambridge Dictionary:\newline
\url{https://dictionary.cambridge.org/dictionary/english/hierarchy}.}. It is a form of organization commonly observed in various  natural, technological, and social complex systems. As other ubiquitous properties such as the small-world property, it is an important characteristic allowing to better understand the entities and the relationship in self-organizing networks. We can  distinguish two types of hierarchies when it comes to network structure:
\renewcommand{\labelenumi}{\roman{enumi}}
\begin{enumerate}
	\item \textbf{Nested hierarchy}: Hierarchy imposed by a system considered on a higher level being composed of subsystems on a lower level that also are composed of subsystems, until we reach the lowest level (can also be called inclusive hierarchy) \cite{e6} .
	\item \textbf{Flow hierarchy}: Hierarchy imposed by the flow of resources that are essential for the system to be maintained. Entities on a higher level influence the ones at a lower level (can also be called control hierarchy) \cite{e6, e8, e9}. 
\end{enumerate}

Table \ref{table:table1} reports the main differences between these two types of hierarchy. In network science literature, hierarchy is often used to quantify the macroscopic network organization \cite{e9, e8, e13, e14, e15, garcia2017ranking, wang2012truss}. In the following, we restrict our attention to the hierarchy measures linked to the microscopic network organization, i.e. at the node level. It can be defined as follows:

\textbf{Hierarchy of nodes}: 
Assume that $G(V,E)$ is an undirected and unweighted graph where $V$ is the set of nodes of size $N=|V|$ nodes and $E \subseteq V \times V$ is the set of edges. The hierarchy of a node $v_{i} \in V$ is given by ${\alpha}(v_i)$. The function $\alpha(v_i)$ is a discrete measure when based on nestedness or mixed hierarchy $(\alpha(v_i) \rightarrow \mathbb{Z}^{+})$ and a real measure when based on flow hierarchy $(\alpha(v_i) \rightarrow \mathbb{R}^{+})$. 

\begin{table}[h]
   \centering
    \caption{Comparing between nested and non-nested hierarchies \cite{e10}.}
    \label{t1}    
      \begin{tabular}{p{7cm}p{7cm}}
      \toprule
    \textbf{Nested Hierarchy} & \textbf{Non-Nested Flow Hierarchy} \\
    \midrule
    Has levels which are composed of and include lower levels & More general as there's a relaxation of containment of lower levels\\
\addlinespace
    Different measurement units at different levels & Same measurement units at different levels \\
    \addlinespace
    Can be applied to directed and undirected networks & Mostly associated with directed networks\\
\addlinespace
    Examples: army consisting of soldiers, class inheritance in object-oriented programming, taxonomies, computer graphic visualization, linguistics, governments, firms and their underlying departments, human body, the universe & Examples: military command (commands flow), food webs (energy flows), pecking orders (dominance flows), software function calling (information flows), demand/supply networks (orders/intermediate goods flow)
 \\
    \bottomrule
       
      \end{tabular}
\label{table:table1}
\end{table}

\subsection{Nested Hierarchy}
Nested hierarchy measures aim at detecting dense structures at various granularity in a network and their hierarchical relations. They are based on the decomposition of the original network resulting in a number of nested entities such that each entity contains or is part of another. The network decomposition forms a hierarchy of induced subgraphs $G^{f}_k(V^{f}_k, E^{f}_k) \subseteq G (V,E)$ where $f$ is the property characterizing the hierarchy and $k$ is its value shared by the nodes.

Core and truss are the two main decomposition methods used to build nested hierarchy. The core decomposition assigns a core number to the nodes based on the degree, while the truss decomposition gives a truss number to the links based on the number of triangles they share. For more information about hierarchical decomposition, one can refer to \cite{malliaros2020core}. 

\subsubsection{K-core Nested Hierarchy}
The $k$-core of a graph $G$  is the maximal subgraph  $G_k^c$ such that every node has at least $k$ neighbors within the subgraph. The $k$-core subgraphs $G_k^c$ are nested. Formally:

$G \supseteq G^c_{k_{min}}$ $\supseteq  G^c_{k_{min+1}} \supseteq ... \supseteq G^c_{k_{min+n}}$ ... $\supseteq$  $G^c_{k_{max}}$
where:
\begin{itemize}
  \item $k_{min}$ is the minimum degree value of the nodes in $G$
  \item $G^c_{k_{max}}$is called the maximal $k$-core subgraph of $G$
\end{itemize}

A node $v_i$ has a core number $c(v_i) = k$, if it belongs to a $k$-core  $G^c_{k}$ but not to the $(k + 1)$-core $G^c_{k+1}$.

The $k$-core subgraphs can be extracted through a peeling process. Starting from the nodes with the minimum degree $k =k_{min}$, nodes that do not have a degree value of $k_{min} + 1$ are removed from the graph and their core number is set to their degree value $k_{min}$. This process is iterated by computing the $k+j$-core. It ends when the maximum  value $k_{max}$ is reached such that the graph $G^c_{k_{max}}$ contains a non-empty $k$-core subgraph. 
The nested structure of $k$-cores reveals a hierarchy by containment. The level of hierarchy of a node $v_i$ is given by:
\begin{equation}
  \alpha_c(v_{i}) = k_{max} - (c(v_i)-1) 
\end{equation}

Figure \ref{fig:fig1} represents a toy example to illustrate the hierarchical decomposition of a network according to the various hierarchy measures introduced in this section and their associated level of hierarchy. The  $k$-core hierarchical decomposition of the given example is reported on the left side of the figure. The minimum degree of the nodes is one, so to extract the $1$-core decomposition of the network $G^c_1$, nodes with a degree smaller than 2 are pruned. A core number $c(v_i)=1$ is assigned to the set of removed nodes $V^c_1=\{2,3,4,12,14,15,16,19\}$. Then all the nodes with a degree value smaller than 3 are removed from the remaining network $G^c_2$. The core number of the set of removed nodes $V^c_2=\{5,6,7,8,9,10,11,13,17,26\}$ is set to $c(v_i)=2$. The process continues by removing from $G^c_3$ all the nodes with a degree value smaller than 4 and assigning to these nodes a core number $c(v_i)=3$. Removing this set $V^c_3=\{1,18,20,21,22,23,24,25\}$ of nodes, the maximum degree value of the original network is reached, and the remaining network $G^c_4$ is empty. 

The hierarchy level ($\alpha_{c}$) of the nodes in the set $V^c_k$ is given by $k_{max} - (j$-core $-1)$. We have $k_{max}=3$, therefore, $V^c_3$ contains the nodes of hierarchy level 1 and so on.

\subsubsection{K-truss Nested Hierarchy}
The truss decomposition is inspired by $k$-core. However, it considers the edges rather than the nodes, and the triangles they participate in. The $k$-truss of a graph $G$ is the maximal subgraph $G_k^t$ such that every link is contained  in at least $k -2$ triangles within the subgraph. The $k$-truss subgraphs $G_k^t$ are nested. Formally:

$G \supseteq G^t_{k_{min}}$ $\supseteq  G^t_{k_{min+1}} \supseteq ... \supseteq G^t_{k_{min+n}}$ ... $\supseteq  G^t_{k_{max}}$
where:
\begin{itemize}
  \item $k_{min}$ is the minimum number of triangles an edge engages in $G$ such that $k_{min} \geq 2$
  \item $G^t_{k_{max}}$ is called the maximal $k$-truss subgraph of $G$
\end{itemize}

An edge $e_{ij}$ has a truss number $t(e_{ij}) = k$, if it belongs to a $k$-truss  $G^c_{k}$ but not to the $(k + 1)$-truss $G^c_{k+1}$.

The $k$-truss hierarchical decomposition can also be obtained through a peeling process. Starting from $k =k_{min}$, edges that are contained in at least $k_{min}$ triangles but not in $k_{min}+1$, are removed and their truss number is set to $k_{min}$. Note that $k_{min}$ starts from 2. The set of resulting isolated nodes $V^t_{k_{min-2}}$ are also removed  and assigned a truss value $k_{min-2}$. The process is iterated by computing the $k_{min}+1$-truss. It ends when the maximum value $k_{max}$ such that the graph $G^t_{k_{max}}$ contains a non-empty $k$-truss subgraph is reached. $k$-truss also reveals a hierarchy by containment. The level of hierarchy of a node $v_i$ is given by:
\begin{equation}
  \alpha_t(v_{i}) = k_{max} - k_{min} - (t(v_i)-1) 
\end{equation}

\begin{figure*}[h]
\begin{center}
\includegraphics[width=1\linewidth, height=5.5 in]{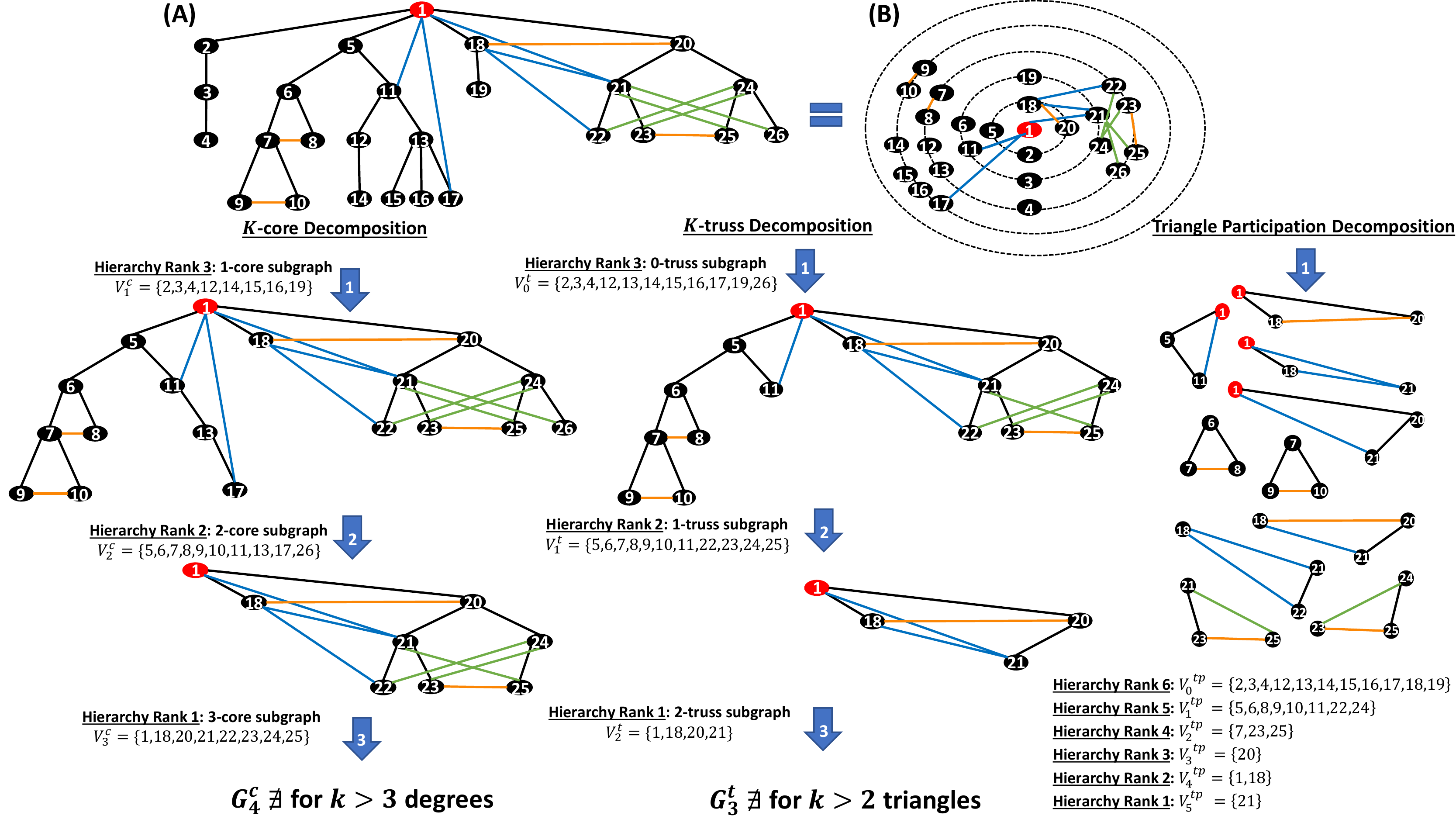}
\end{center}
\caption{Obtaining hierarchy ranking of nodes for graph $G$ with $N$=26 and $E$=38. (A) Example of Tree-based view of $G$ with irregularities. (B) Radial-based view of $G$ with irregularities. The irregularities are shown in different colors. The blue links indicate "shortcut" connections. The orange links indicate in-layer connections. The green links indicate fully connected trees at a specific rank. Hierarchical decomposition based on k-core (on the left) and k-truss (in the middle) and triangle participation (on the right) are illustrated.}
\label{fig:fig1}
\end{figure*}

In the example of figure \ref{fig:fig1}, the $k$-truss hierarchical decomposition is shown in the middle. Some edges are not engaging in any triangle, $k_{min}$ is set to 2. This results in the $0$-truss decomposition of the network $G^t_0$, where edges which do not participate in any triangle are removed. Accordingly, an edge truss number $t(e_{ij})=0$ is assigned to removed edges. The set of nodes isolated  by the removal of the edges are also removed and assigned the same truss number $V^t_0=\{2,3,4,12,13,14,15,16,17,19,26\}$. Then, $k$ is incremented to 3. All edges that engage in one triangle $(3 - k_{min})$ but not in two triangles are removed. They are assigned an edge trussness $t_e(e_{ij})=1$. The set of associated isolated nodes are also removed $V^t_1=\{5,6,7,8,9,10,11,22,23,24,25\}$ resulting in the $1$-truss network $G^t_1$.
The process continues by removing all edges that participate in two triangles and not three triangles are removed. The set of removed edges  are assigned a truss number $t(e_{ij})=2 $ together with the set of nodes isolated by this operation $V^t_2=\{1,18,20,21\}$. The maximum trussness is reached at $G^t_2$ because there is no edge participating in more than two triangles. The truss values range from 0 to 2.

The hierarchy level ($\alpha_{t}$) of the nodes in the set $V^t_k$ is given by $k_{max} - k_{min} - (j$-truss $- 1)$. We have $k_{min}=2$, $k_{max}=4$ and $t(v_i)=2$, therefore, $V^c_2$ contains the nodes of hierarchy level 1 and so on.

\subsection{Flow Hierarchy}

Many real-world networks such as information networks and production networks are better characterized by the flows of resources rather than a containment ordering. Indeed, flows are essential for the systems to produce, reproduce and transform. In such systems the entities are organized into a flow of hierarchy. Therefore, flows determine different types of interactions, ascertaining different levels of importance among nodes. A hierarchy measure based on flow assigns a real hierarchy value to each node $(\alpha(v_i) \rightarrow \mathbb{R}^{+})$. The hierarchical level is based on this value.


Indeed, if $|\alpha(v_i)|<|\alpha(v_{j})| : \forall v_{i},v_{j} \in V$ then node $v_{j}$ is more important (higher hierarchically) than node $v_{i}$.

There have been a handful of work on flow hierarchy \cite{e8, e13, e14}, but most of the work quantifies the hierarchy of the whole graph, which is not our focus. To our knowledge, Local Reaching Centrality (LRC) is one of the flow hierarchy measures quantifying the hierarchy of nodes.

\subsubsection*{Local Reaching Centrality Flow Hierarchy}
Flow hierarchy measures the influence of a node on all other nodes in a network and that is captured by local reaching centrality (LRC) \cite{e9}.

LRC was originally developed for directed and weighted graphs, considered as a generalization of the $m$-reach centrality that takes into consideration nodes that are within $m$ distance of a given node. Authors in \cite{e9} consider ($m=N$) where $N$ is the number of nodes in $G$ and define LRC as follows:
\begin{equation}
\alpha_l(v_i)=\frac{1}{(N-1)} \sum_{{j=1}}^{{N}} \left( \frac{\sum_{k=1}^{d^{out(v_i,v_j)}} \omega_{v_i}^{(k)}(v_j)}{d^{out}(v_i,v_j)} \right)
\end{equation}
where:
\begin{itemize}
  \item ${d^{out}(v_i,v_j)} $ is the length of the directed path that goes from node $v_i$ to $v_j$ via an outgoing edges such that ${d^{out}(v_i,v_j)} < \infty$
  \item $\omega_{v_i}^k$ is the weight of the $k$-th edge on its given path
  \item $N$ is the total number of nodes in the network
\end{itemize}

However, LRC can be used for unweighted and undirected graphs. In this case, it reduces to closeness centrality for disconnected graphs:
\begin{equation}
\alpha_l(v_i)=\frac{1}{(N-1)}\sum_{j=1}^{{N}}\frac{1}{d(v_i,v_j)}
\end{equation}
where:
\begin{itemize}
  \item $d(v_i,v_j)$ is the distance between nodes $v_i$ and $v_j$ such that $d(v_i, v_j)<\infty$
  \item $N$ is the total number of nodes in the network
\end{itemize}

From equation 3 and 4, we can see that the hierarchy is a continuous number between [0,1]. The higher the value, the more the node is able to impact other nodes, and the higher it is in terms of hierarchy. 

\subsection{Mixed Hierarchy}
Inspired by transitivity red based on triplets that quantifies the extent to which nodes in a network tend to form dense clusters (also called the ratio of transitive triplets), we propose to consider  "\textit{Triangle Participation}" as a mixed hierarchy measure for nodes \cite{Newman2010}. Indeed, it has the essence of both nested and flow hierarchy. Note that this measure is generally not used to quantify hierarchy but rather as a ratio in community detection problems \cite{harenberg2014community, yang2015defining, mothe2017community, Wagenseller2016CommunityDA}. This proposition is based on four main reasons:

\begin{enumerate}
	\item It can be considered as a flow hierarchy measure for each node as it conveys flow of information probability according to how many times a node participates in a triangle (closed triad). The higher the number of triangles' participation, the more the node is able to diffuse and acquire information.
	
	\item Simultaneously, it can be viewed as a nested hierarchy measure for each node as it allows to categorize the nodes according to the number of triangles they participate in. The hierarchy level of a node is therefore based on the density of the motifs that emerge as for $k$-core and $k$-truss.
	
	\item  Due to the irregularities found in trees (resulting in triangles, as shown in figure \ref{fig:fig1} where most triangles have at least one irregularity), $k$-core assumes only dyadic relationships, thus is myopic of modeling the irregularities happening at different levels in a hierarchy. On the other hand, even though $k$-truss is capable of quantifying number of triangles, and it may be too coarse as it allows to have minimum $k-2$ triangles participation.
	
	\item  $k$-core and $k$-truss are based on a more complex extraction process that may be prohibitive on large scale networks.
\end{enumerate}

It would be interesting to design a measure  which relaxes the complexity of the extraction process of nested hierarchy and at the same time is able to capture irregularities. Triangle participation overcomes these challenges. The hierarchy measure $\alpha(v_i) \rightarrow \mathbb{Z}^{+}$ divides the nodes $V$ into groups $V^f_k \in G$ where $f$ is the property function defined on the network and $k$ is the property value shared by the nodes. Groups are not subgraphs since they are not based on higher levels containing lower levels as $k$-core and $k$-truss. If nodes $v_{i}, v_{j}$ share the same property value $k$, then they share the same mixed hierarchy level $\alpha(v_i) \rightarrow \mathbb{Z}^{+}$. Hence, both nodes belong to the same group (set) where $v_{i}, v_{j} \in V^f_k$.

\subsubsection*{Triangle Participation Mixed Hierarchy}
To represent the mixed hierarchy measure, we need to first go on a couple of definitions. Considering graph $G(V,E)$ :
\begin{itemize}
  \item  Let the adjacency matrix $A = (a_{i,j})$ describe the connectivity of the graph $G$ such that: 
        \begin{equation}
        a_{i,j}=
        \begin{cases}
            1,& \text{if node $v_i$ is connected to node $v_j$ } \\
            0,              & \text{otherwise}
            
        \end{cases}
        \end{equation}
   \item Let the neighborhood of any node $v_i$ be defined as the set $\mathcal{N}_p(v_i)={\{v_j \in V: (v_i,v_j) \in E\}}$ at length $p$, where $p=1,2,...,D$. $D$ is the diameter of $G$. Accordingly, two nodes are neighbors of order $A^p$ if there's a minimal path connecting them at $p$ steps.
  
\end{itemize}

Let nodes $u, v, w$ be three nodes forming a closed triangle $\Delta_{uvw}$. Triangle participation of a node $v_{i}$ is simply the number of triangles it is in. It is defined as:
\begin{equation}
Tp(v_{i}) = \sum |\Delta_{uvw}|
\end{equation}
where $|\Delta_{uvw}|$ is the number of times node $v_i$ exists in triangle $\Delta_{uvw}$, $\forall  u,v,w \in V$. The level of hierarchy of a node $v_i$ is given by:
\begin{equation}
\alpha_{tp}(v_{i}) = k_{max} + 1 - Tp(v_i)
\end{equation}

Referring to the example in figure \ref{fig:fig1}, the triangle participation hierarchical computation is shown on the right. Starting from $k_{min}=0$, that is nodes not participating in any triangle, we obtain the first group of nodes $V^{tp}_0=\{2,3,4,12,13,14,15,16,17,18,19\}$. $k_{min}$ is now incremented by 1, obtaining the second group of nodes, which participate in 1 triangle only $V^{tp}_1=\{5,6,8,9,10,11,22,24\}$. Nodes in one group do not belong to another group as the $k_{min}$ value is the highest number of triangles they are in. The process continues until we reach the highest number of triangles $k_{max}=5$ with $V^{tp}_5=\{21\}$. As it can be seen, there is no a priori in this approach. The number of triangles for each node is counted only once (there's only one blue downward flash, unlike with $k$-core and $k$-truss decomposition). Moreover, there are more levels in mixed hierarchy. In other words, more heterogeneity among the nodes. 

The hierarchy level ($\alpha_{tp}$) of the nodes in the set $V^{tp}_k$ is given by $ k_{max} + 1 - Tp(v_i)$. There is $k_{max}=5$ triangles, therefore, $V^{tp}_5$ contains the nodes of hierarchy level 1 and so on.

\section{Centrality}
\label{sec:Centrality}

Identifying the most influential nodes in a network using centrality measures is one of the main research issues in network science. As the notion of "importance" is subject to various interpretations, there is a great deal of work on the subject. For a complete overview about the various measures, the reader can refer to the surveys \cite{survey1, survey2}. The vast majority of centrality measures can be classified as neighborhood-based, path-based or iterative refinement-based measures \cite{lu2016vital}. While these measures are usually linked to a single topological property of the network, recent works turn to multidimensional definitions. In this case, various complementary scalar topological properties of the network are combined to quantify the influence of the nodes \cite{multidimensional, ghalmane2019immunization,ibnoulouafi2018m,gupta2016centrality}.Complexity is also an important issue of centrality measurement. Measures can be classified as local or global depending of the information used. Global measures assume the knowledge of the overall network to compute the centrality of a node, while local measures need only information in the neighborhood of the node \cite{das2014local,ghalmane2018betweenness}. Of course, global measures are generally more effective, but at the expense of a higher complexity that can be prohibitive, especially for large networks \cite{protein}. 

In order to explore the relations between centrality and hierarchy we choose to restrict our attention to the most influential centrality measures. Based on the taxonomy adopted in \cite{lu2016vital}, they belong to the three main groupings: neighborhood-based, path-based, and iterative refinement-based centrality measures.

\subsection{\textbf{Neighborhood-Based Centrality}}

Neighborhood-based centrality quantifies the importance of a node according to the influence it can exert on its local surroundings.  Those local and semi-local measures are generally easy to compute, but they are totally agnostic about the overall network structure.

\subsubsection{Degree Centrality}
The degree centrality of a node is one of the simplest centrality measures. It is proportional to the number of neighbors directly linked to the node. The more connections, the higher its influence on  neighboring nodes. It requires low computation for identifying important nodes. It is defined as follows:
\begin{equation}
\beta_d(v_i)=\frac{1}{(N-1)}\sum_{j=1}^{N}a_{ij}
\end{equation}
where:
\begin{itemize}
  \item $a_{ij}$ is obtained from $A^{1}$, 1-step neighborhood ($p$=1)
  \item $N$ is the total number of nodes in the network
\end{itemize}

\subsubsection{Local Centrality}
Local centrality extends degree centrality by increasing the size of the neighborhood of a node. While degree centrality consider direct neighbors, local centrality takes into consideration not only direct neighbors, but the neighbors of the neighbors. This is because the direct neighborhood alone may not be fully informative regarding a node's importance. It is defined as follows:
\begin{equation}
\beta_l(v_i)=\frac{1}{(N-1)}\sum_{j=1}^{N} a_{ij}^{\ensuremath{'}}
\end{equation}
where:
\begin{itemize}
  \item $a_{ij}^{\ensuremath{'}}$ is obtained from $A^{2}$, 2-step  neighborhood ($p$=2)
  \item $N$ is the total number of nodes in the network
\end{itemize}

\subsection{\textbf{Path-Based Centrality}}
 Path-based centrality measures quantify the ability of a node to spread information throughout the entire network. As they consider the paths going through each node, those global measures are difficult to compute in large-scale networks.
\subsubsection{Betweenness Centrality}
Betweenness centrality quantifies the importance of a node based on the fraction of shortest paths between any two nodes that pass through it. It is defined as follows:
\begin{equation}
\beta_b(v_i)=\frac{2}{(N-1)(N-2)}\left(\sum_{s,t\neq i }{\frac{\sigma_{v_i}(v_s,v_t)}{\sigma(v_s,v_t)} }\right)
\end{equation}
where:
\begin{itemize}
  \item $\sigma(v_s,v_t)$ is the number of shortest paths between nodes $v_s$ and $v_t$
  \item $\sigma_{v_i}(v_s,v_t)$ is the number of shortest paths between nodes $v_s$ and $v_t$ that pass through node $v_i$ 
  \item $N$ is the total number of nodes in the network
\end{itemize}

\subsubsection{Current-Flow Closeness Centrality}
Current-flow closeness centrality (also called information centrality) quantifies the node's importance according to the information transmitted along paths \cite{brandes2005centrality}. It is defined as follows:
\begin{equation}
 \beta_c(v_i)= \frac{N}{\sum_{j=1} r_{ii} +r_{jj}-2r_{ij}}
\end{equation}
where:
\begin{itemize}
  \item $r_{ij}$ is the  amount of information that can be transmitted from node $v_i$ to $v_j$ throughout all possible paths
  \item $N$ is the total number of nodes in the network
\end{itemize}
$r_{ij}$ is an element of the matrix $R$ defined as follows: 
$R=[D-A+F]^{-1}$ where  $D$ is a $N-$dimensional diagonal matrix with the degree of the nodes along its diagonal and 0 everywhere else,
$A$ is the adjacency matrix of the network,
and $F$ is a matrix with all its elements equal to 1.

\subsection{\textbf{Iterative Refinement-Based Centrality}}
As path-based centrality measures, iterative refinement centrality measures make use of the topology of the overall network structure. Therefore they are global centrality measures. However, in this case, a node importance is linked to the importance of each of its neighbors. 

\subsubsection{Katz Centrality}
Katz centrality quantifies the importance of a node in such a way that it takes into consideration the influence of all nodes and their paths with respect to it. However, as nodes become more distant from the node under study, their influence is attenuated.
It is defined as follows:
\begin{equation}
\beta_k(v_i)= \sum_{p=1} \sum_{j=1} s^p a^p_{ij}
\end{equation}
where:
\begin{itemize}
  \item $a^p_{ij}$ is the connectivity of node $v_i$ with respect to all the other nodes at $A^p$
  \item $s^p$ is the attenuation parameter where $s \in$ [0,1]
\end{itemize}
As the distance between nodes $p$ increases, the attenuation factor $s^p$ decreases the influence of the other nodes $v_j$ connected to $v_i$. Note that the attenuation parameter $s$ should be strictly less than the inverse of the largest eigenvalue ($\lambda_{max}$) of the adjacency matrix $A$ to have a solution for Katz centrality.

\subsubsection{PageRank Centrality}
PageRank centrality works in a similar way that Katz centrality. It also takes into consideration the quantity and quality of nodes. However, PageRank is based on the probability of random walks. The adjacency matrix $A$ is transformed to a stochastic matrix $P$ representing the probability of visiting a node. Then, the centrality works iteratively using the power iteration till reaching a steady state. Originally, it is  defined on directed graphs. The undirected version is defined as follows:
\begin{equation}
\beta_p(v_i)=\frac{1-d}{N} + d \sum_{v_j \in M_i} \frac{\beta_p(v_j)}{k_j}
\end{equation}
where:
\begin{itemize}
  \item $\beta_p(v_i)$ is the PageRank centrality of node $v_i$
  \item $\beta_p(v_j)$ is the PageRank centrality of node $v_j$
  \item $M_i$ is the set of nodes linked to node $v_i$
  \item $k_j$ is the number of links from node $v_j$ to node $v_i$
  \item $d$ is the damping parameter where $d \in$ [0,1] ensuring convergence in case $k_j=0$
  \item $N$ is the total number of nodes in the network
\end{itemize}
Note that the damping parameter value $d$ is fixed at 0.85 in the subsequent experiments.

\section{Evaluation measures}
\label{sec:evalmeasures}
In this section the various correlation and similarity measures used to investigate the relationship between hierarchy and centrality measures are presented. In addition to these measures, the $k$-means algorithm and the Schulze voting method used for deeper investigation are briefly presented. A detailed description is given in the section \ref{sec:DataAndMethodology} about the utilization of these measures.

\subsection{Correlation}
Three correlation measures are used in order to compare the set of hierarchy values with the set of centrality values for a given triplet (network, hierarchy measure, centrality measure).  Pearson correlation is based on the values to compare, while Spearman correlation and Kendall Tau are rank-based comparisons.

\subsubsection{Pearson Correlation}
Pearson's correlation coefficient is a popular measurement for the linear strength and direction between two variables. Its value ranges between [-1, +1]. The value -1 indicates a high negative correlation while the value +1 indicates a high positive correlation. A value of 0 means that there is no correlation at all. 

Assume that the set of hierarchy measures $\alpha_i$ and that the set of centrality measures $\beta_i$ for a network of size $N$ is given, the Pearson's correlation coefficient between the two measures is computed as follows:

\begin{equation}
    \rho_p(\alpha,\beta) = \frac{\sum_{i=1}^{N}(\alpha_i - \bar{\alpha})(\beta_i - \bar{\beta} )}{\sqrt{\sum_{i=1}^{N} (\alpha_i - \bar{\alpha} )^{2} \sum_{i=1}^{N}(\beta_i - \bar{\beta})^{2}}}
\end{equation}
where:
\begin{itemize}
  \item$\bar{\alpha}=\frac{\sum_i^N \alpha_i}{N}$ is the average of the hierarchy measure values
\item $\bar{\beta}=\frac{\sum_i^N \beta_i}{N}$ is the average of the centrality measure values
  \item $N$ is the number of nodes in the network
\end{itemize}

\subsubsection{Spearman Correlation}
Spearman's correlation coefficient is a modified version of Pearson's. Considering the ranks of the variables instead of their raw value, it measures their monotonic relationship. Monotonic relationships are less restrictive than linear relationships in case there is large variance but a relationship between variables still exists. Spearman's correlation  values range also between [-1, +1]. 
Assume that the set of hierarchy measures $\alpha_i$ and that the set of centrality measures $\beta_i$ for a network of size $N$ is given, the Spearman correlation coefficient between the two measures is:

\begin{equation}
\rho_s(\alpha,\beta) = \frac{\sum_{i=1}^{N}(R(\alpha_i) - \overline{R(\alpha)})(R(\beta_i) - \overline{R(\beta)})}{\sqrt{\sum_{i=1}^{N} (R(\alpha_i) - \overline{R(\alpha)} )^{2} \sum_{i=1}^{N}(R(\beta_i) - \overline{R(\beta)})^{2}}}
\end{equation}

where:
\begin{itemize}
  \item $R(\alpha_i)$ and $R(\beta_i)$ represent the rank of the $i$-th hierarchy measure and centrality measures respectively
  \item  $\overline{R(\alpha)}$ and $\overline{R(\beta)}$ are the average value of the ranks of the hierarchy and centrality measures respectively
  \item $N$ is the number of nodes in the network
\end{itemize}

\subsubsection{Kendall Tau's Correlation}
Kendall Tau's correlation is a modified version of Spearman's. It also considers ranks and its values ranges between [-1, 1]. Furthermore, it takes into consideration the order of ranks as well. Suppose that $R(\alpha)$ and $R(\beta)$ are the ranking lists of the hierarchy measure and centrality measure, respectively. Kendall Tau's correlation determines the strength of the ordinal association based on the concordance (ordered in a same manner) and discordance (ordered differently) between the pairs in the ranking lists $R(\alpha)$ and $R(\beta)$. A pair of nodes $v_i$ and $v_j$ is concordant if $R(\alpha(v_i)) > R(\alpha(v_j))$ and $R(\beta(v_i)) > R(\beta(v_j))$ or if $R(\alpha(v_i)) < R(\alpha(v_j))$ and $R(\beta(v_i)) < R(\beta(v_j))$. While the pair is discordant if $R(\alpha(v_i))) > R(\alpha(v_j))$ and $R(\beta(v_i)) < R(\beta(v_j))$ or if $R(\alpha(v_i)) < R(\alpha(v_j))$ and $R(\beta(v_i)) > R(\beta(v_j))$. Kendall Tau's version that takes into consideration ties as well is used. It is denoted by $\tau_b$, where the pair is tied if $R(\alpha(v_i)) = R(\alpha(v_j))$ and/or $R(\beta(v_i)) = R(\beta(v_j))$. The Kendall Tau's correlation coefficient is defined as follows:
\begin{equation}
    \tau_b(\alpha,\beta) = \frac{n_c-n_d}{\sqrt{(n_c+n_d+x)(n_c+n_d+y)}}
\end{equation}
where:
\begin{itemize}
  \item $n_c$ is the number of concordant pairs
  \item $n_d$ is the number of disconcordant pairs
  \item $x$ is number of tied pairs on the hierarchy $\alpha$ variable
  \item $y$ is number of tied pairs on the centrality $\beta$ variable
  
\end{itemize}

\subsection{Similarity}
Two similarity measures are used in order to compare the set of hierarchy values with the set of centrality values for a given triplet (network, hierarchy measure, centrality measure).  Jaccard similarity is used to measure the proportion of common nodes in the top-k values in the two sets, while Rank-Biased Overlap (RBO) allows to compare the top-k values and also the entire sets. 
\subsubsection{Jaccard's Similarity}
The Jaccard's similarity index measures the similarity between two finite sets of data. Suppose that $\alpha$ and $\beta$ are two sets containing the labels of a group of nodes extracted using a hierarchy measure and a centrality measure, respectively from a given network. The Jaccard index is defined as the size of the intersection divided by the size of the union of the sample sets. Formally it is given by:
\begin{equation}
    J(\alpha, \beta) = \frac{|\alpha \cap \beta|}{|\alpha \cup \beta|}
\end{equation}

The Jaccard's similarity index values range between [0,1]. If there is no common nodes in the two sets $J=0$, and $J=1$ if all the members of the first set exist in the second set.

\subsubsection{Rank-Biased Overlap Similarity}
Rank-Biased Overlap (RBO)\cite{webber2010similarity} measures the similarity of two ordered sets. It is based on Jaccard's similarity index but with indefinite sets at specific depth $d$. It allows to give more weight to differences at the top of the ranked sets than differences further down. It is a similarity measure on the full rankings which is consistent whatever the depth of the evaluation. In other words, its value increases if the agreement between the two sets increases with deeper evaluation and it decreases if the agreement goes down.

Assuming $\alpha$ and $\beta$ are two infinite ordered sets of the hierarchy and centrality measures respectively, the RBO between those two sets is defined as follows:
\begin{equation}
    RBO(\alpha,\beta) = (1-p){\sum_{d=1}^{\infty} p^{(d-1)}} \frac{|\alpha_{d} \cap \beta_{d}|}{d}
\end{equation}
where:
\begin{itemize}
    \item $p$ is the probability of continuing to the next rank, while $(1-p)$ is the probability of stopping at a specific rank
  \item $d$ is the depth reached on sets $\alpha,\beta$ from position 1
    \item $|\alpha_{d} \cap \beta_{d}|/{d}$ is the proportion of the similarity overlap between hierarchy and centrality sets at depth $d$
\end{itemize}
RBO values range between [0,1]. There is no similarity between the two ranked sets if its value is null. A value of 1 indicates that the two ranked sets are identical.

\subsection{K-means Clustering Algorithm}
$k$-means is an unsupervised clustering algorithm that categorizes objects based on their feature values. The $k$ parameter specifies the number of clusters, it is an input of the algorithm. Given a set of $m$ samples ($m_1$, $m_2$, ..., $m_o$) where each sample is composed of a $d$-dimensional feature vector, $k$-means divides the $m$ observations into disjoint clusters $C = \{C_1, C_2, ..., C_k\}$  by minimizing the within-cluster criterion: 
\begin{equation}
    \sum_{i=1}^{o}\min_{\mu_j \in C}(||m_i - \mu_j||^2)
\end{equation}
where:
\begin{itemize}
  \item $m_i$ is the $d$-dimensional feature vector of sample $i$
  \item $\mu_j$ is the mean of the samples $m$ in cluster $C$
\end{itemize}

The $k$-means algorithm is used to cluster networks according to the various evaluation measures (correlation and similarity) across all possible combinations of the hierarchy and centrality measures used.

\subsection{The Schulze Method}

 The Schulze method \cite{schulze2011new} is a voting scheme that gives a single-winner or a sorted list of winners according to votes as indicated by user preferences. It transforms the lists of ordered preferences (ties being allowed) into a matrix of pairwise preferences $\Omega$. From this matrix, another matrix is extracted, expressing the strength between pairwise preferences $\Upsilon$. Strength extraction is defined on directed paths. The weakest link from all possible strongest paths between two candidates, determines the strength of the winning between one candidate to another. The Schulze method is given in Algorithm 1. It uses a variant of the Floyd-Warshall algorithm to compute the strength of the strongest paths.

The Schulze voting scheme is used to rank the strength of occurrence among the hierarchy and centrality combinations. The ranking depends on the combinations' correlation and similarity magnitude across the networks used.

\begin{algorithm}
\caption{Schulze Voting Method}
\begin{algorithmic}
\REQUIRE $C$ candidates list and $\Omega(i,j)$ matrix consisting of the number of voters who prefer candidate $i$ over candidate $j$ 

\textbf{// Step 1: Initialization of Strength Matrix}
\FOR{$i,j=0$ in $C$ where $i \neq j$}
            \IF{$\Omega(i,j) > \Omega(j,i) $}
                \STATE $\Upsilon(i,j)  \leftarrow \Omega(i,j)$
            \ELSE
                \STATE $\Upsilon(i,j)  \leftarrow 0$
            \ENDIF
\ENDFOR

\textbf{// Step 2: Strongest Paths Calculation}
\FOR{$i,j=0$ in $C$ where $i \neq j$}
        \FOR{$k=0$ in $C$}
            \IF{$i \neq k$ and $j \neq k$}
                \STATE $\Upsilon(j,k)  \leftarrow  max [\Upsilon(j,k), min(\Upsilon(j,i), \Upsilon(i,k))]$
            \ENDIF
        \ENDFOR
\ENDFOR

\textbf{// Step 3: Ranking Final Set}
\FOR{$i,j=0$ in $C$ where $i \neq j$}
    \IF{$\Upsilon(i,j) > \Upsilon(j,i) $}
        \STATE $L[i] \leftarrow L[i] + 1 $
    \ENDIF
      \STATE $sort(L[i])$
\ENDFOR

\RETURN $L$ 

\end{algorithmic}
\end{algorithm}

\begin{table*}[ht!]
 \caption{Basic topological properties of real-world networks under investigation. $N$ is the network size. $|E|$ is the number of edges. $k_{min}, k_{max}$ are the minimum and maximum degree, respectively. $<k>$ is the average degree. $<d>$ is the average shortest path. $\nu$ is the density. $\zeta$ is the transitivity (also called global clustering coefficient). $k_{nn}(k)$ is the assortativity (also called degree correlation coefficient). $\gamma_{max}$ is the maximum coreness. $\varphi_{max}$ is the maximum trussness. * indicates the topological properties of the largest connected component of the network in case it is disconnected.}
\label{table:characteristics}
\begin{tabular}{cccccccccccc}

\toprule
     &  $N$ & $|E|$ & $k_{min}, k_{max}$ & $<k>$ & $<d>$ & $\nu$ & $\zeta$  & $k_{nn}(k)$ & $\gamma_{max}$  & $\varphi_{max}$ \\
\midrule
\multicolumn{12}{c}{\textbf{Animal Social Networks}}\\
\midrule
Mammals &  28      & 235      & [4, 23]  & 16.78   & 1.34    & 0.716   &  0.727    & -0.004    & 14  & 12   \\ 

Insects &  113      & 4,550      & [42, 109]  & 80.53   & 1.28    &  0.798  & 0.785     & -0.030    & 60   & 45  \\ 

Birds*   & 117      & 304          & [1, 21]   & 5.19  &  4.36   & 0.577   & 0.472    & 0.062     &  8   &  9    \\ 

Reptiles*   &  496        & 984        & [1, 17]    & 3.96   & 8.08    & 0.008  & 0.419    & 0.342     & 8   & 9  \\

\midrule
\multicolumn{12}{c}{\textbf{Biological Networks}}\\
\midrule
Mouse Visual Cortex &  193        & 214        & [1, 31]    & 2.21   & 4.22   & 0.011   & 0.004    & -0.844     & 2   & 3     \\ 

E.Coli Transcription* &  329        & 456        & [1, 72]    &  2.77 &  4.84   & 0.008    & 0.023    & -0.263     & 3   & 3     \\ 

Yeast Protein* &  1,458        & 1,993        & [1, 56]    & 2.73   & 6.77    & 0.001   & 0.051    & -0.207     & 5   & 6     \\ 

Human Protein* &   2,217      & 6,418        & [1, 314]    & 5.78   & 3.84    & 0.002   & 0.007    & -0.331    & 10   & 5     \\ 

\midrule
\multicolumn{12}{c}{\textbf{Human Social Networks}}\\
\midrule
Zachary Karate Club &  34        & 78      & [1, 17]    & 4.58   &  2.36    & 0.139   &  0.255   &  -0.475    & 4   & 5     \\ 

Madrid Train Bombings &  64        & 243        & [1, 29]    & 7.59   & 2.69    & 0.120   & 0.561    & 0.029     & 10   & 11     \\

Physicians* &  117        & 465        & [2, 26]    & 7.94   & 2.58    & 0.068   & 0.174    & -0.084     & 6   & 5     \\

Adolescent Health &  	2,539        & 12,969        & [1, 36]    & 10.21   & 4.55    & 0.002   & 0.141    & 0.231     & 7   & 7    \\

\midrule
\multicolumn{12}{c}{\textbf{Miscellaneous Networks}}\\
\midrule

Les Mis\'erables &  77        & 254        & [1, 36]    & 6.59   & 2.63    & 0.086   & 0.498    & -0.165     & 9   & 10      \\ 

World Metal Trade* &  80        & 875        & [4, 77]    & 21.62   & 1.72    & 0.276   & 0.459    & -0.391     & 14   & 14     \\ 

Adjective Noun &  112       & 425        & [1, 49]    & 7.58   & 2.53    & 0.068  & 0.156    & -0.129     & 6   & 5     \\

Internet Autonomous Systems &  6,474        & 12,572        & [1, 1458]    & 3.88   & 3.68    & 0.0006   & 0.009    & -0.181     & 12   & 10     \\
\midrule
\multicolumn{12}{c}{\textbf{Infrastructure Networks}}\\
\midrule
U.S. States &  49        & 107        & [1, 8]    & 4.36   & 4.16    & 0.090   & 0.406    & 0.233     & 3   & 3 \\

U.S. Airports &  500        & 2,980        & [1, 145]    & 11.92   & 3.01    & 0.023   & 0.351    & -0.267     & 29   & 27     \\ 

EuroRoads* &  1,039        & 1,305        & [1, 10]    & 2.51   & 18.31    & 0.002   & 0.035    & 0.090     & 2   & 3     \\ 

U.S. Power Grid &  4,941        & 6,594        & [1, 19]   & 2.66   & 18.98    & 0.0005   & 0.103    & 0.003     & 5   & 6     \\ 

\midrule
\multicolumn{12}{c}{\textbf{Collaboration Networks}}\\
\midrule
NetScience Collaboration &  379        & 914        & [1, 34]    & 4.82   & 6.04    & 0.012   & 0.430    & -0.081     & 8   & 9     \\ 

CS Ph.D. Collaboration* &  1,025        &  1,043        & [1, 46]    & 2.03   & 11.55   & 0.001   & 0.002    & -0.253     & 2   & 3     \\ 

GrQc Collaboration &  4,158        & 13,422        & [1, 81]    & 6.45   & 6.06    & 0.001   & 0.628   & 0.639     & 43   & 44     \\ 

AstroPh Collaboration* &  17,903        & 196,972        & [1, 504]    & 22.004   & 4.19    & 0.001   & 0.317    &  0.201 & 56   & 57     \\ 
\midrule
\multicolumn{12}{c}{\textbf{Online Social Networks}}\\
\midrule

Retweets Copenhagen &  761        & 1,029       &    [1, 37] & 2.70   & 5.30    & 0.003   & 0.060    & -0.099     & 4   & 4     \\ 

Facebook Ego &  4,039        & 88,234        & [1, 1045]    & 43.69   & 3.69    & 0.010   & 0.519    & 0.063     & 115   & 97     \\ 

Facebook Politician Pages &  5,908        & 41,729        & [1, 323]    & 14.12   & 4.69    & 0.002   & 0.301    & 0.018     & 31   & 26     \\

PGP-based Social Network &  10,680        & 24,316        & [1, 205]    & 4.55   & 7.57    & 0.0004   & 0.378    & 0.238     & 31   & 27     \\ 

\bottomrule

\end{tabular}
\end{table*}

\section{Data and Methods}
\label{sec:DataAndMethodology}
This section presents briefly the data used in the experiments and the experimental process. Note that unweighted and undirected networks are used. In case the original network is made of multiple components, only the largest connected component is retained.

\subsection{Data}
To investigate extensively the relationship between hierarchy and centrality, 28 real-world networks originating from various domains such as social, biological, ecological, infrastructure networks have been selected. Their sizes range from tens to thousands of nodes. A brief description of these networks is provided. Table \ref{table:characteristics} reports their basic topological properties. Note that all the data sets are available online \cite{nr, latora2017complex, de2018exploratory, kunegis2014handbook}.

\subsubsection{Animal Social Networks}
\begin{itemize}
    \item \textbf{Mammals (mammalia-sheep-dominance):} The nodes represent sheep. They are connected if they have a dominance fight against each other. Interactions convey the need for superiority within a group \cite{nr}.
    
      \item \textbf{Insects (insecta-ant-colony1-day01):} The nodes represent ants. They are connected in case they are enclosed within the same trapezoidal regions. Interactions convey group membership\cite{nr}.
    \item \textbf{Birds (aves-weaver-social):}  The nodes represent weaver birds. They are connected if they use the same bird nest for roosting or building within a year. Interactions convey social projection  bipartite \cite{nr}.
    \item \textbf{Reptiles (reptilia-tortoise-network-fi):} The nodes represent tortoises. They are connected if they use the same hole in the ground for refuge. Interactions convey social projection bipartite \cite{nr}.

\end{itemize}

\subsubsection{Biological Networks}
\begin{itemize}
    \item \textbf{Mouse Visual Cortex (bn-mouse-visual-cortex-2):} Nodes are neurons in the visual cortex of the brain that is responsible for processing visual information. Interactions represent fiber tracts that connect one neuron to another \cite{nr}.

    \item \textbf{E.Coli Transcription:} Nodes are Escherichia Coli bacteria regulating the conversion of DNA to RNA, responding to various biological signals. Interactions represent transcriptions and regulations of genes \cite{latora2017complex}.
        
    \item \textbf{Yeast Protein (bio-yeast-protein-inter):} A protein-protein interaction network where a protein is connected to another in case a direct interchange takes place. Interactions represent chemical reactions \cite{nr}.
    
        \item \textbf{Human Protein (maayan-figeys):} A protein-protein interaction of human cells, obtained from a first large-scale study on humans. Interactions represent chemical reactions \cite{kunegis2014handbook}.

\end{itemize}

\subsubsection{Human Social Networks}
\begin{itemize}

    \item \textbf{Zachary Karate Club:} Members of a karate club in a university are connected to each other in case they interact with each other outside the club. Interactions represent friendship \cite{latora2017complex}.
    \item \textbf{Madrid Train Bombings:} Nodes are terrorists of the Madrid train bombing on March 11, 2004. Interactions represent contact among two terrorists \cite{kunegis2014handbook}.
    \item \textbf{Physicians:} Nodes are physicians in U.S. towns. Interactions between two physicians represent trust. There is a link between two physicians if one of them asks for advice from another, wants to discuss a given topic, or is his friend \cite{kunegis2014handbook}.
        
    \item \textbf{Adolescent Health:} Nodes are students asked to list 5 of their best female friends and 5 of their best male friends. Interactions represent friendship ties \cite{kunegis2014handbook}.

\end{itemize}

\subsubsection{Miscellaneous Networks}
\begin{itemize}
    \item \textbf{Les Mis\'erables:} Actors in Victor Hugo's novel `Les Mis\'erables' connected to each other if they appeared in the same chapter of the novel. Interactions represent co-appearances \cite{latora2017complex}.
        
    \item \textbf{World Metal Trade (world\_trade):} World metal trade in 1994. Nodes represent countries involving heavy metal or high-technology metal products manufactures. Interactions represent trades from one country to another \cite{de2018exploratory}.
    
    \item \textbf{Adjective Noun (adjnoun):} Nodes are the most common occurring adjectives and nouns in "David Copperfield" novel of Charles Dickens. Interactions represent adjacent positions of any pair of words in the novel \cite{nr}.
        
    \item \textbf{Internet Autonomous Systems (AS-20000102):} Nodes are autonomous systems (AS). Interactions represent connections for exchanging information between two AS. The network is a snapshot of the Internet on the 2$^{nd}$ of January, 2000 \cite{latora2017complex}.

    \end{itemize}

\subsubsection{Infrastructure Networks}
\begin{itemize}
    \item \textbf{U.S. States (contiguous-usa):} Nodes are the states of America. An edge represents border sharing between any two states. Hawaii and Alaska are excluded because they aren't adjacent to the rest of the states \cite{nr}.
    
    \item \textbf{U.S. Airports:} Nodes represent the airports in America. They are connected if there is a direct flight between two corresponding airports. Interactions represent direct transport connection among the airports \cite{kunegis2014handbook}.
    
    \item \textbf{EuroRoads (inf-euroroad):} Nodes are European cities. Roads across the European continent are the links connecting cities that may be within the same country or not. Interactions represent direct transport connection among cities \cite{nr}.
        
    \item \textbf{U.S. Power Grid:} Nodes are either a generator, transformer, or substation in the western states of America. Interactions between nodes represent a power supply line \cite{kunegis2014handbook}.

\end{itemize}

\subsubsection{Collaboration Networks}
\begin{itemize}
    \item \textbf{NetScience Collaboration (ca-netscience):} Nodes are researchers in Network Science. Interactions represent co-authorship of scientific papers \cite{nr}.

    \item \textbf{GrQc Collaboration (ca-GrQc):} Nodes are researchers co-authoring in General Relativity and Quantum Cosmology. Interactions represent co-authorship of scientific papers \cite{nr}.
    
    \item \textbf{CS Ph.D. Collaboration (ca-CSphd):}  Nodes are  Ph.D. students and their supervisors specializing in Computer Science. Interactions represent collaboration of passing scientific knowledge \cite{nr}. 
    
    \item \textbf{AstroPh Collaboration (ca-AstroPh):}  Nodes are researchers co-authoring in Astrophysics, obtained from e-print arXiv. Interactions represent co-authorship of scientific papers \cite{nr}.

\end{itemize}

\subsubsection{Online Social Networks}
\begin{itemize}

    \item \textbf{Facebook Ego Network (ego-facebook):} Nodes are users on Facebook, collected by a survey of participants using the Facebook application. Interactions represent online friendship \cite{nr}.

    \item \textbf{Twitter Retweets Copenhagen (rt-twitter-copen):} Nodes are Twitter users with their retweets gathered over a alongside a United Nations conference in Copenhagen about climate change. Interactions represent retweets \cite{nr}.   
    
    \item \textbf{Facebook Politician Pages (fb-pages-politician):} Nodes represent politician verified pages on Facebook from different countries. Interactions represent  mutual likes among the politicians \cite{nr}.

     \item \textbf{PGP-based Social Network:} Nodes are users of the web of trust, sharing information under the Pretty Good Privacy (PGP) algorithm. Interactions represent mutual secure information sharing among users \cite{kunegis2014handbook}.

\end{itemize}

\begin{figure}[ht!]
\begin{center}
\includegraphics[width=0.65\linewidth, height= 1.5 in]{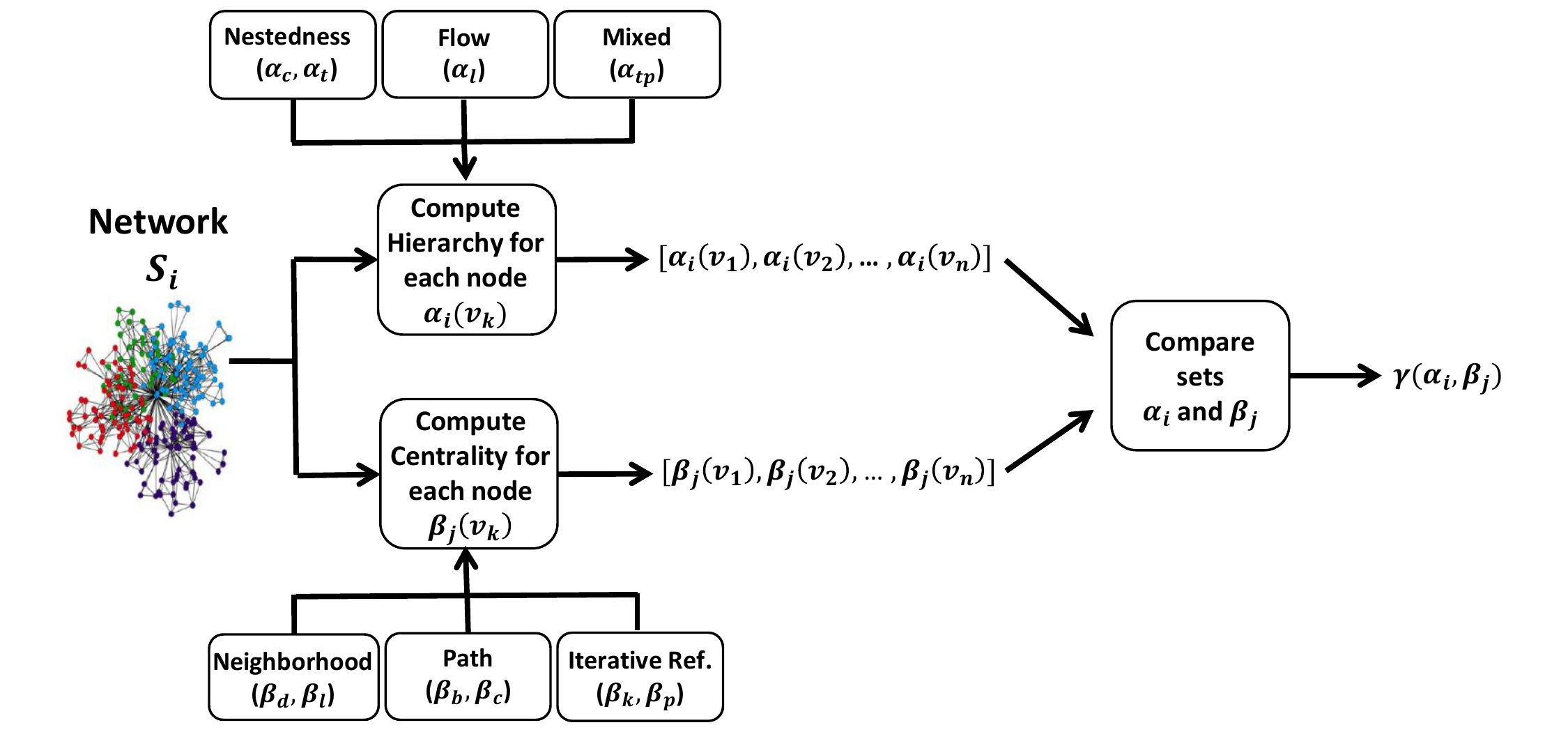}
\end{center}
\caption{Comparing a hierarchy $ \alpha_i$ and centrality $ \beta_j$ measure using an evaluation measure $ \gamma(\alpha_i,\beta_j)$ for a given network $S_i$. $k$-core $ \alpha_c $ and $k$-truss $ \alpha_t$ are nested hierarchy measures. LRC $ \alpha_l$ is a flow hierarchy measure.  Triangle participation $ \alpha_{tp}$ is a mixed hierarchy measure. Degree $ \beta_d $, and local $\beta_l$ are neighborhood-based centrality measures. Betweenness $ \beta_b$ and Current-flow closeness $ \beta_c$ are path-based centrality measures. Katz $\beta_k$ and PageRank $ \beta_p$ are iterative refinement-based centrality measures. The evaluation measure $ \gamma(\alpha_i,\beta_j)$ can be a correlation or a similarity measure.}
\label{fig:exp1}
\end{figure}

\subsection{Methods}
A series of experiments are conducted in order to characterize the relations between the set of hierarchy measures $A =\{\alpha_c,\alpha_t, \alpha_l,  \alpha_{tp}\}$ and the set of centrality measures $B=\{\beta_d,\beta_l,\beta_b,\beta_c, \beta_k, \beta_p\}$  using a set of real-world networks $ S =\{S_1,S_2, ...,S_{10},..., S_{28}\}$, a set of correlation evaluation measures $\rho =\{\rho_p,\rho_s,\tau_b\}$ and a set of similarity evaluation measures $Sim=\{J,RBO\}$. In the following study, for the four hierarchy measures $\alpha_i$ $(i=1\rightarrow4)$ used, notation is as following: $\alpha_c$ is $\alpha_1$, $\alpha_t$ is $\alpha_2$, $\alpha_l$ is $\alpha_3$, and $\alpha_{tp}$ is $\alpha_4$. For the six centrality measures $\beta_j$ $(j=1\rightarrow6)$ used, notation is as following: $\beta_d$ is $\beta_1$, $\beta_l$ is $\beta_2$, $\beta_b$ is $\beta_3$, $\beta_c$ is $\beta_4$, $\beta_k$ is $\beta_5$, and $\beta_p$ is $\beta_6$.

\subsubsection{Comparing the various combinations of centrality and hierarchy measures for each network}
In the first set of experiments, the  aim is to compare the hierarchy measures to the centrality measures two-by-two for a given network. These experiments allow us to answer the main question of this study, that is, do hierarchy measures convey complementary information as compared to centrality measures given the topology of the network? Figure \ref{fig:exp1} illustrates the experimental process. In order to compare a hierarchy measure $\alpha_i$ to a centrality measure $\beta_j$ for a given network $S_i$ of size $N$, the sample set of hierarchy $\{\alpha_i(v_1),\alpha_i(v_2),...,\alpha_i(v_l),...,\alpha_i(v_n)\}$ and the sample set of the centrality $\{\beta_j(v_1),\beta_j(v_2),...,\beta_j(v_l),...,\beta_j(v_n)\}$ measures are computed for all the nodes of the network. Then a comparison measure $\gamma (\alpha, \beta)$ is computed. This process is performed for the 28 networks under investigation using all the centrality (Degree, Local, Current-Flow Closeness, Betweenness, Katz, and PageRank) and hierarchy ($k$-core, $k$-truss, LRC, and triangle participation) measures two-by-two. So, for each network, 24 combinations of hierarchy and centrality measures are evaluated using Pearson, Spearman and Kendall Tau correlation measures. 
Results of these experiments allow to investigate if there are significant patterns that appear in terms of correlation. Additionally, the consistency of the pairwise correlation measures across networks can be evaluated. Finally, one can check if the results given by the various correlation measures are consistent.

Similarly, hierarchy measures are compared with centrality measures two-by-two using the two similarity measures (Jaccard and RBO) for a given network. Jaccard index checks the similarity across two finite sets regardless of rank. The sample set of centrality and the sample set of the hierarchy measures are computed for the top 10 nodes in small networks ($N$<150). For bigger networks, the top 10\% are considered. RBO is designed for infinite sets taking into consideration the ranking of the nodes and their ties. Furthermore, it allows to assign a higher weight to top nodes using a tuning parameter $p$. Two values of the tuning parameter are used ($p$=0.5 and $p$=0.9). In order to compare with the results based on the Jaccard index, the similarity of the top 10 nodes for small networks or the top 10\% for bigger networks are computed using RBO. Additionally, comparison are performed considering all the nodes (the entire set of nodes) of the network for either small or big networks. Hence, there is 4 versions of RBO (at different $p$ and on top-k and on entire set of nodes). Results of these experiments allow to check the consistency across the similarity measures. Furthermore, comparisons can be performed with the results obtained using the correlation measures. 

\begin{figure}[ht!]
\begin{center}
\includegraphics[width=0.65\linewidth, height= 1.5 in]{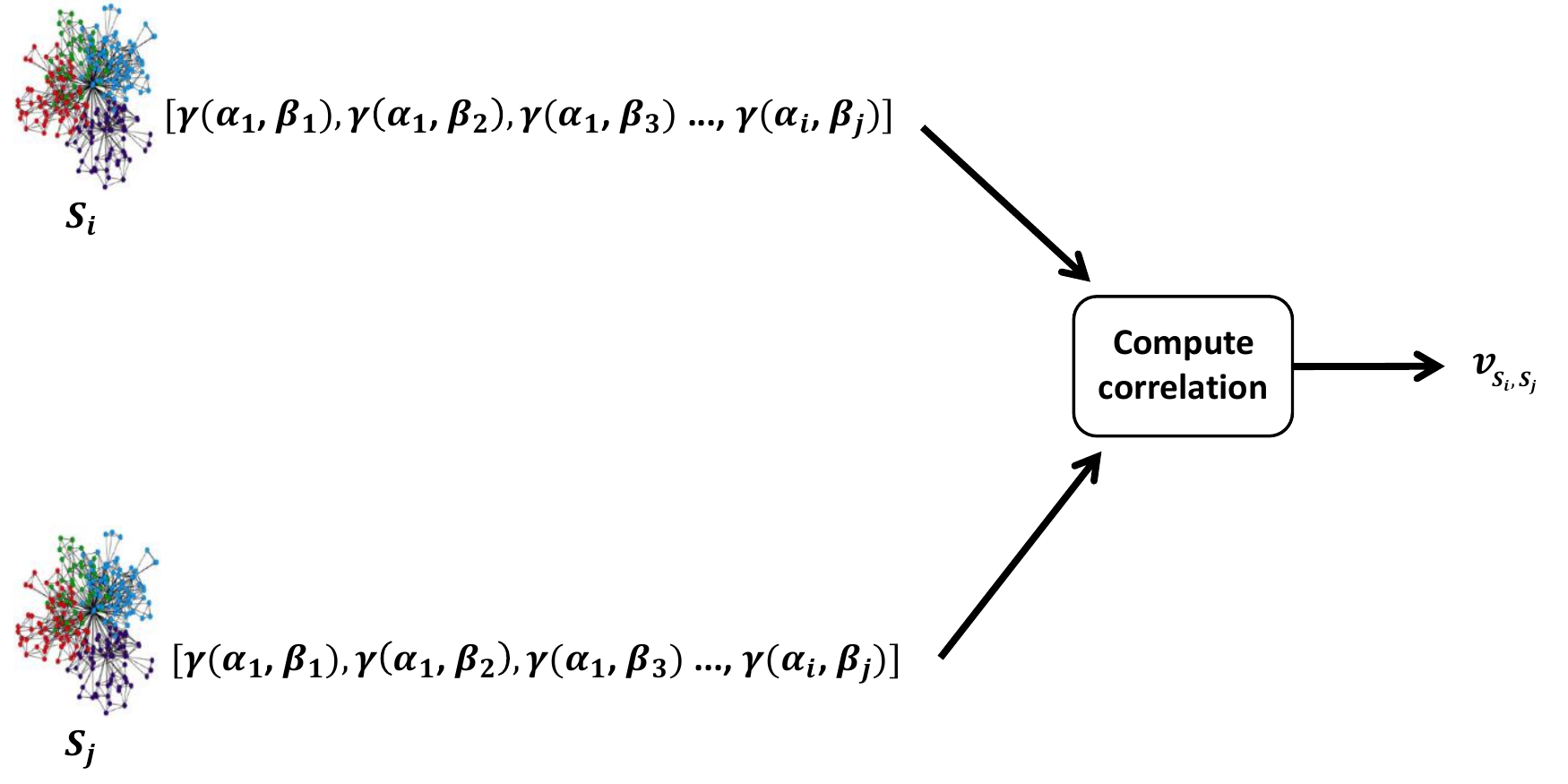}
\end{center}
\caption{Comparing two networks $S_i, S_j$ in terms of correlation after the computation of their hierarchy $ \alpha_i$ and centrality $ \beta_j$ measure using an evaluation measure  $ \gamma(\alpha_i,\beta_j)$. $a_i$ is a nested hierarchy measure ($k$-core $ \alpha_c $ and $k$-truss $ \alpha_t$ ) or a flow hierarchy measure (LRC) $ \alpha_l$ or a mixed hierarchy measure (Triangle participation $ \alpha_{tp}$). $\beta_j$ is a  neighborhood-based centrality measure (Degree $ \beta_d $, and local $\beta_l$) or a  path-based centrality measure (Betweenness $ \beta_b$ and Current-flow closeness $ \beta_c$) or an iterative refinement-based centrality measures (Katz $ \beta_k$ and PageRank $ \beta_p$). The evaluation measure $ \gamma(\alpha_i,\beta_j)$ can be a correlation or a similarity measure.}
\label{fig:expCorCor}
\end{figure}

\subsubsection{Comparing the networks according to the evaluation measures sample sets}

The second set of experiments is based on the previous results. The goal is to deepen the understanding of the interplay between centrality, hierarchy, and network topology. In other words, the following questions are raised: are there clusters of networks that can be discovered based on the correlation between centrality and hierarchy measures? That is, do networks show similar behavior based on their values of correlation and similarity between hierarchy and centrality combinations? And do the networks in these clusters share some specific topological properties? This investigation is based on three experiments.

In the first experiment, the goal is to check if the networks exhibit a similar behavior based on the various centrality and hierarchy combination. For a given evaluation measure $\gamma$, a network $S_i$ is associated to the sample set $\Gamma_i= \{\gamma(\alpha_1,\beta_1), \gamma(\alpha_1,\beta_2),..., \gamma(\alpha_i,\beta_j)\}$. For all the pairs of networks $S_i, S_j$, the Pearson correlation between their sample sets $\rho_p(\Gamma_i , \Gamma_j)$ is computed. Only Pearson correlation is used because, in this case, ranking does not matter. This experiment is performed for all the evaluation measures under test. Figure \ref{fig:expCorCor} illustrates the experimental process. The corresponding correlation matrices are represented using a heatmap.

\begin{figure}[ht!]
\begin{center}
\includegraphics[width=0.65\linewidth, height= 1.5 in]{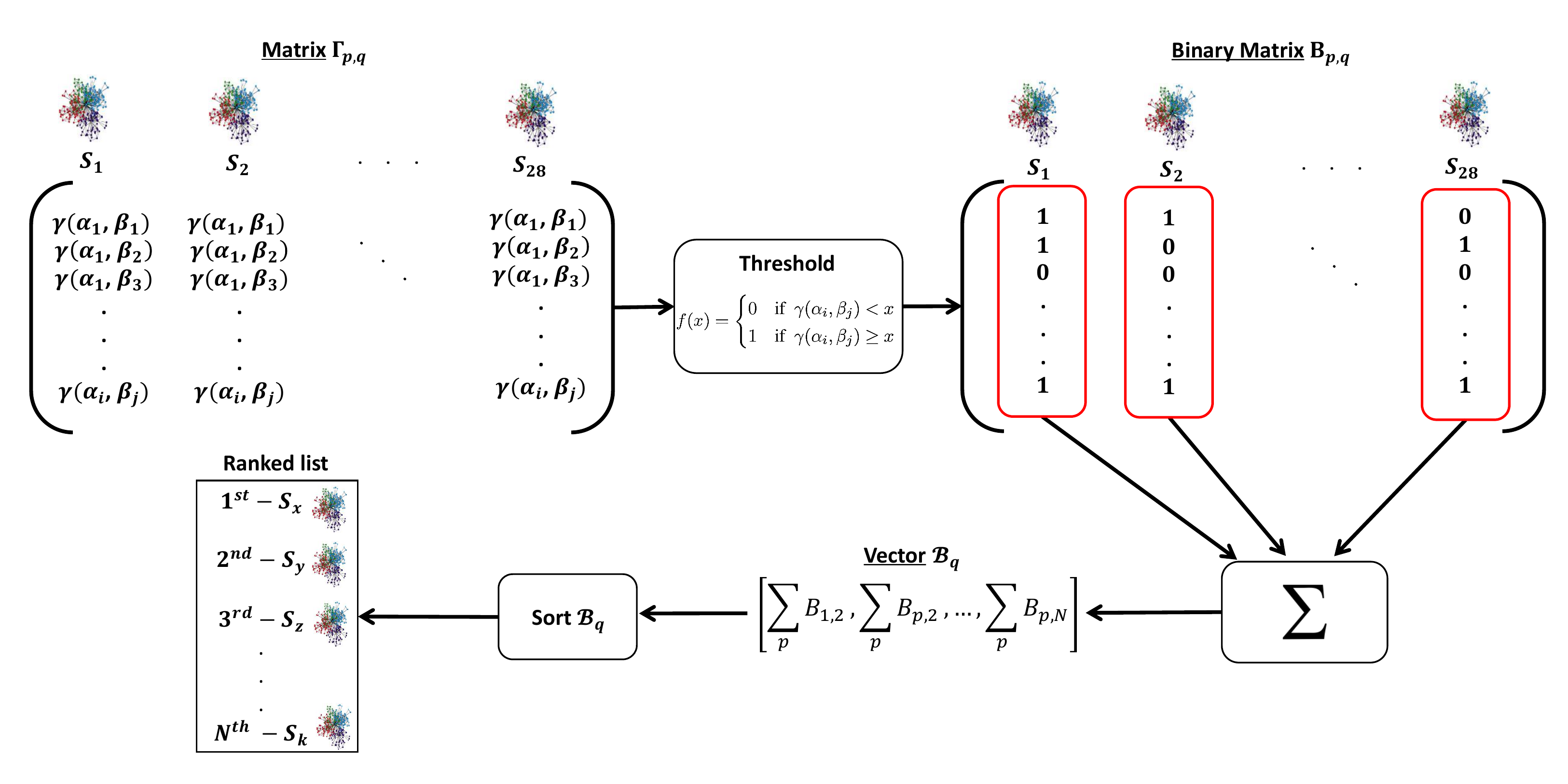}
\end{center}
\caption{Ranking all networks $ S =\{ S_1,S_2, ...,S_{10},..., S_{28}\}$ in descending order after comparing the values of their evaluation measure $\gamma(\alpha_i,\beta_j)$ to a threshold  $f(x)$. Starting from matrix $\Gamma_{p,q}$ a binary matrix $B_{p,q}$ is formed.Summing all the elements of its columns, the final ranking for each network is obtained. $a_i$ is a nested hierarchy measure, and it includes $k$-core $ \alpha_c $ and $k$-truss. $ \alpha_t$. LRC $ \alpha_l$ as flow hierarchy measure. Triangle participation $ \alpha_{tp}$ is a mixed hierarchy measure. $\beta_j$ is a neighborhood-based centrality measure, and it includes Degree $ \beta_d $, and Local centrality $ \beta_l$. Betweenness $ \beta_b$ and Current-flow closeness $ \beta_c$ are path-based centrality measures. Katz $ \beta_k$ and PageRank $ \beta_p$ are iterative refinement-based centrality measures. The evaluation measure $ \gamma(\alpha_i,\beta_j)$ can be either a correlation or a similarity measure.}
\label{fig:expThresh} 
\end{figure}

The second experiment aims to extract common topological characteristics from the networks based on their ranking. In this case, the evaluation measure between a centrality measure and a hierarchy measure is binarized. More precisely, $\gamma(\alpha_i,\beta_j)$ is set to one if the evaluation measure is above a threshold value $\mu$ and set to zero otherwise. Then, a score is assigned to each network according to the number of times the evaluation measure of the various combination between centrality and hierarchy has been assigned a value of one. Finally, the networks are ranked according to this score. 

Figure \ref{fig:expThresh} illustrates this process.
The same process is used for all the evaluation measures. In all the experiments the  threshold value $ \mu$ is set to $0.7$. Indeed, it is generally admitted as a high value for similarity and correlation.

\begin{figure}[ht!]
\begin{center}
\includegraphics[width=0.65\linewidth, height= 1.5 in]{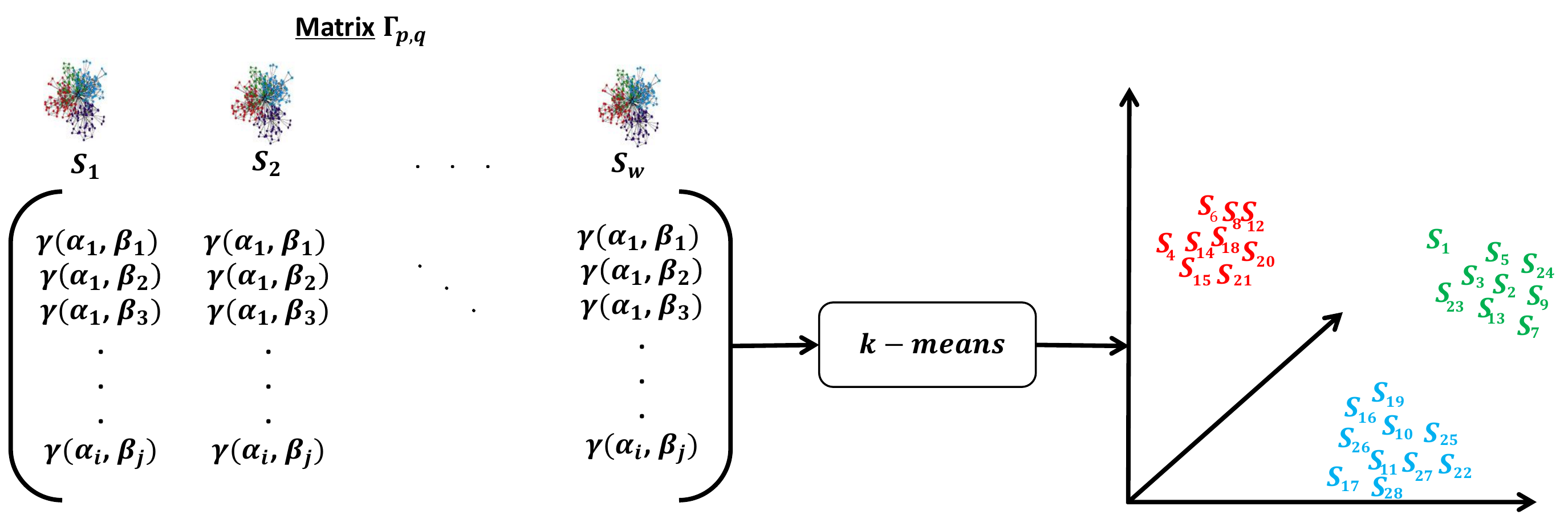}
\end{center}
\caption{Clustering all networks $ S =\{ S_1,S_2, ...,S_{10},..., S_{28}\}$ using $k$-means algorithm according to their evaluation measure values $\gamma(\alpha_i,\beta_j)$ as features set. $a_i$ is a nested hierarchy measures ($k$-core $ \alpha_c $ and $k$-truss $ \alpha_t$) or a flow hierarchy measure (LRC $ \alpha_l$) or a mixed hierarchy measure (Triangle participation $ \alpha_{tp}$).  $\beta_j$ is a neighborhood-based centrality (Degree $ \beta_d $, and local $ \beta_l$) or a path-based centrality measures (Betweenness $ \beta_b$ and Current-flow closeness $ \beta_c$) or an iterative refinement centrality measures Katz ($ \beta_k$ and PageRank $ \beta_p$). $ \gamma(\alpha_i,\beta_j)$) can be a correlation or a similarity measure.}
\label{fig:kmeans} 
\end{figure}

Finally, the third experiment aims at categorizing the sample set of networks $S_i$ according to their multidimensional feature set $\Gamma_i= \{\gamma(\alpha_1,\beta_1), \gamma(\alpha_1,\beta_2),..., \gamma(\alpha_i,\beta_j)\}$. The process based on the  $k$-means algorithm is depicted in figure \ref{fig:kmeans}.

\subsubsection{Comparing the combinations of hierarchy and centrality measures using the Schulze voting method}

\begin{figure}[ht!]
\begin{center}
\includegraphics[width=0.65\linewidth, height= 1.5 in]{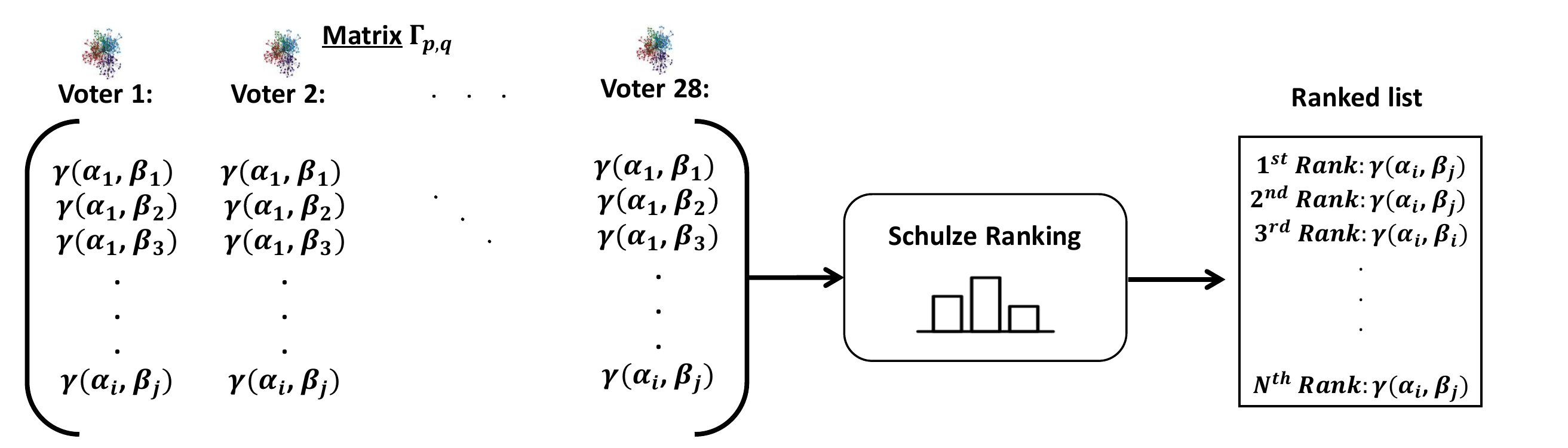}
\end{center}
\caption{Ranking all possible combinations between hierarchy $\alpha_i$ and centrality $\beta_j$ based on their evaluation measure values $\gamma(\alpha_i,\beta_j)$ using the Schulze Method. Networks are the voters  and the evaluation of the combinations of the various centrality and hierarchy measures $\gamma(\alpha_i,\beta_j)$ are the candidates. $a_i$ is a nested hierarchy measure ($k$-core $ \alpha_c $ and $k$-truss $ \alpha_t$) or a flow hierarchy measure (LRC $ \alpha_l$) or a as mixed hierarchy measure (Triangle participation $ \alpha_{tp}$ ). $\beta_j$ is a neighborhood-based centrality measure (Degree $ \beta_d $, and local $ \beta_l$) or a path-based centrality measure (Betweenness $ \beta_b$ and Current-flow closeness $ \beta_c$) or an iterative refinement-based centrality measures (Katz $ \beta_k$ and PageRank $ \beta_p$). The evaluation measures $ \gamma(\alpha_i,\beta_j)$ can be a correlation or a similarity measure.}
\label{fig:expShulze}
\end{figure}

The third set of experiments aims to answer the question: what are the hierarchy measures which are the most distant from the centrality measures independently of the network topology. In order to explore this question the Schulze voting method is used to rank the different combinations. The 28 networks $S_i$ are considered as the voters.  Given an evaluation measure  $\gamma $ the 24 combinations between hierarchy and centrality measures are the candidates. Each network expresses a set of preferences about the 24 candidate combinations of centrality and hierarchy based on the magnitude of a evaluation measure $\Gamma_i= \{\gamma(\alpha_1,\beta_1), \gamma(\alpha_1,\beta_2),..., \gamma(\alpha_i,\beta_j)\}$. The couple of centrality and hierarchy measures are then ranked according to the preference of all the voters. Note that, if a specific combination with high preference occurs frequently among the networks, it is highly ranked. Additionally, if a combination with lower preference also occurs frequently, it is also highly ranked, but at a lower rank that the one with higher preference  and comparable frequency. The Schulze analysis is performed for all the correlation and similarity measures. A general view of the Schulze method using networks as voters is given in figure \ref{fig:expShulze}.

\section{Experimental Results}
\label{sec:ExperimentalResults}
In this section, we report the results of the empirical evaluation about the relations between hierarchy, centrality and network topology  based on the several experiments conducted. In order to improve reading fluency typical results are presented in this section, and complementary results are reported in supplementary materials.

\subsection{Comparing the various combinations of centrality and hierarchy measures for each network}
In this series of experiments, the evaluation measures (3 correlation measures and 5 similarity measures) between the 6 centrality measures and the 4 hierarchy measures have been computed for each of the 28 networks. Heatmaps are used to present the results.

\begin{figure*}[h!]
\begin{center}
\includegraphics[width=1\linewidth, height=2.6 in]{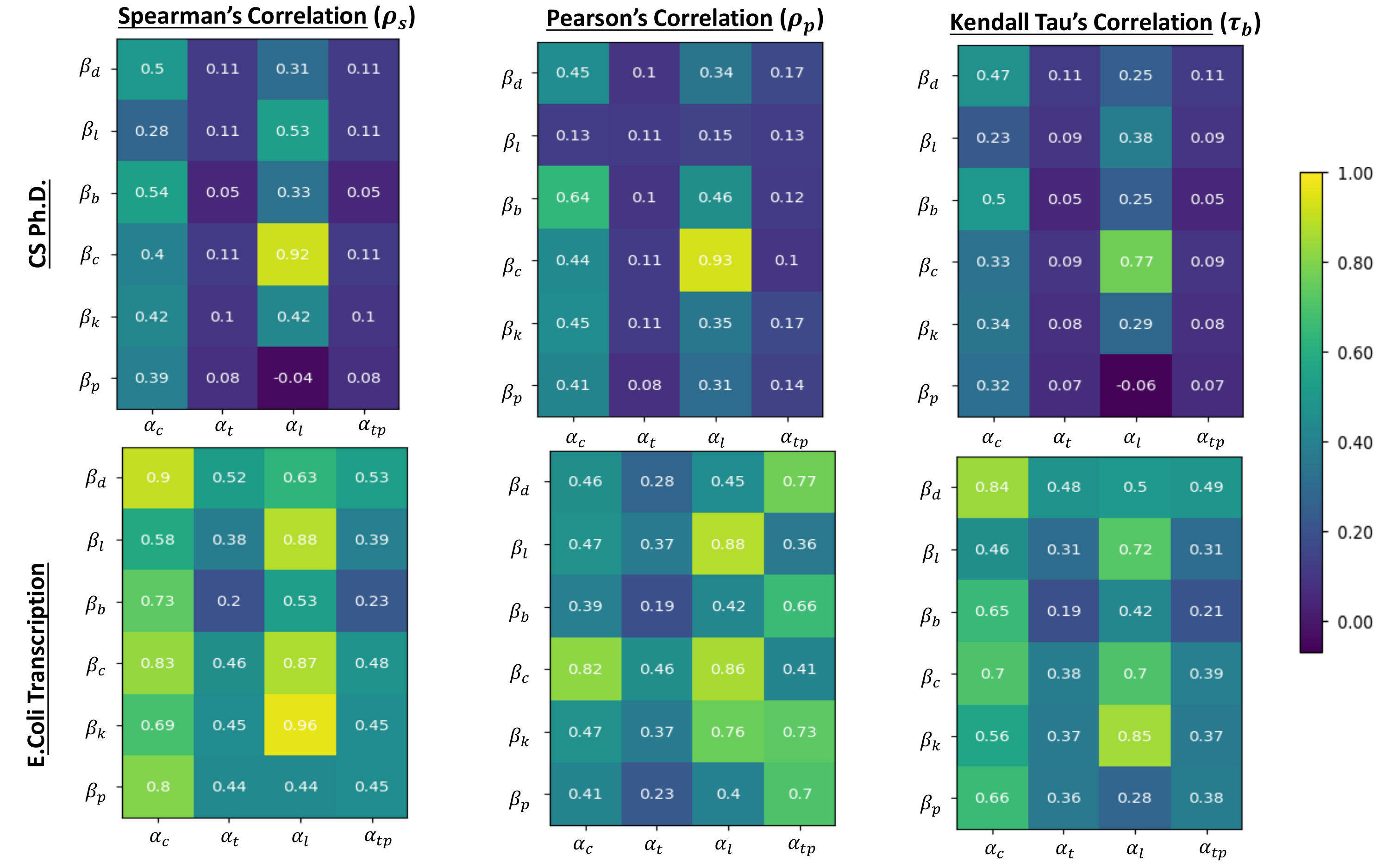}
\end{center}

\end{figure*}

\begin{figure*}[ht!]
\begin{center}
\includegraphics[width=1\linewidth, height=2.6 in]{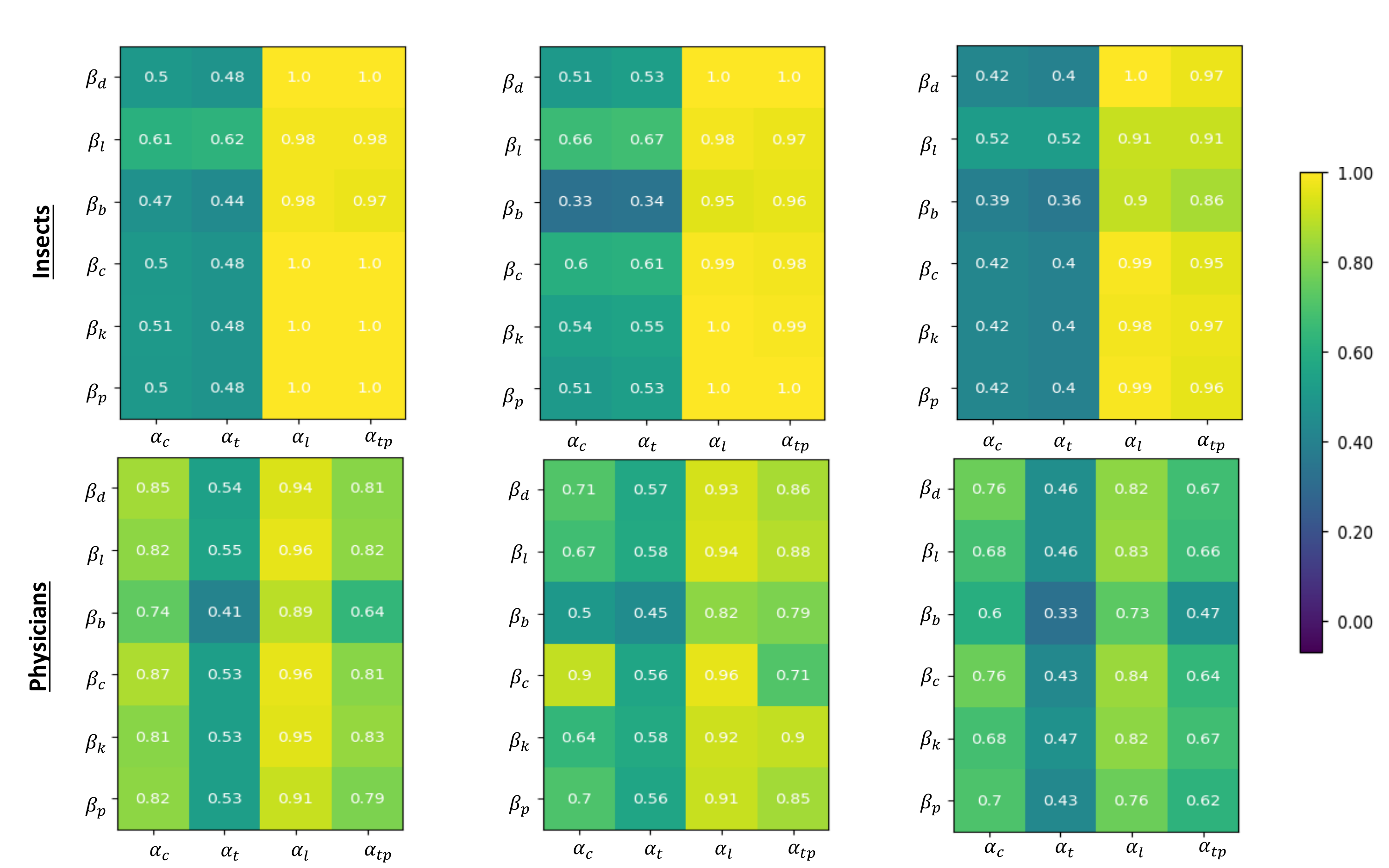}
\end{center}
\end{figure*}

\begin{figure*}[ht!]
\begin{center}
\includegraphics[width=1\linewidth, height=2.6 in]{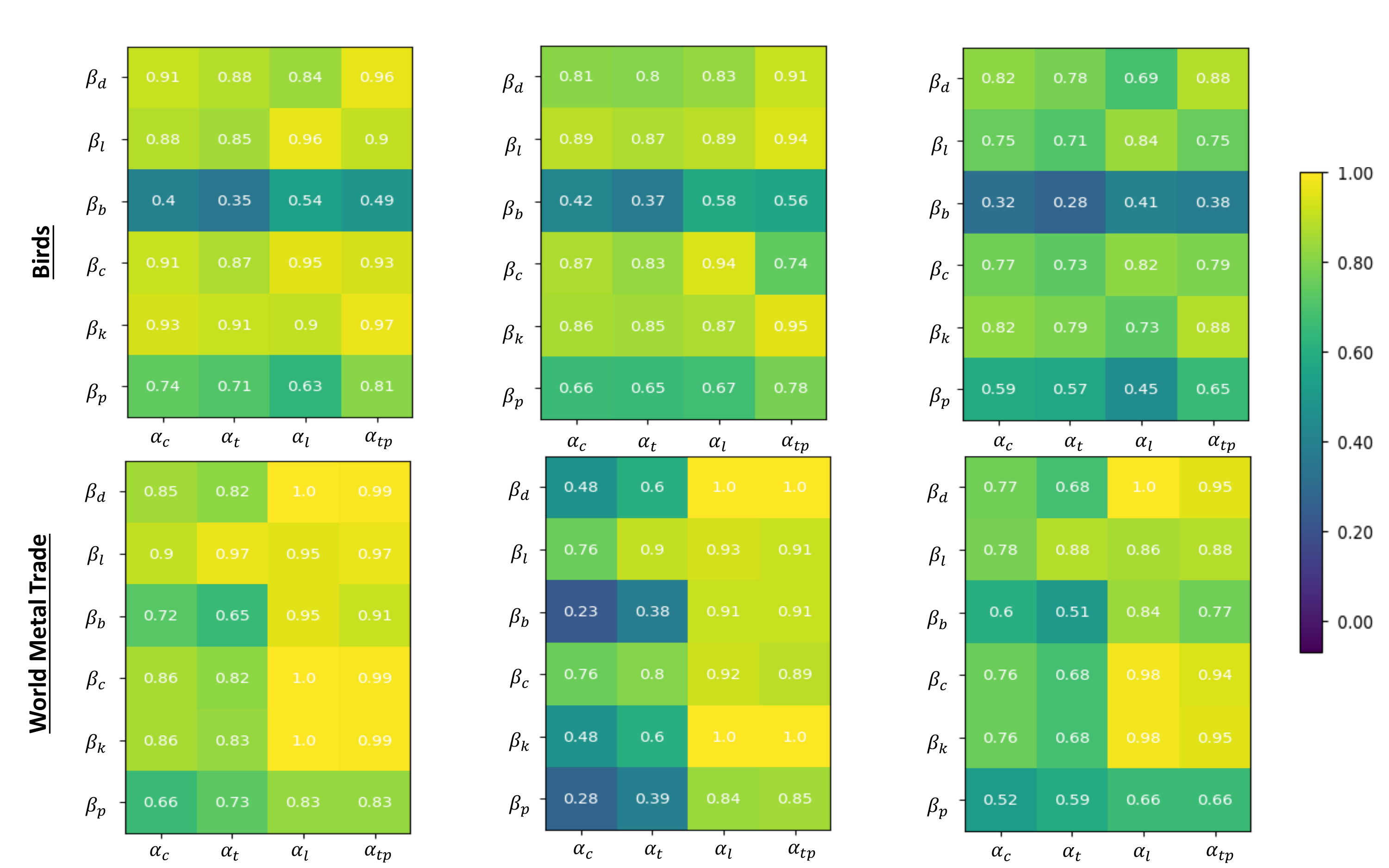}
\end{center}
\caption{Heatmaps of the correlation evaluation measures for the various combinations of hierarchy $\alpha_i$ and centrality $\beta_j$  measures of six real-world networks (from top to bottom). The hierarchy measures are $\alpha_{c}$ = $k$-core, $\alpha_{t}$ = $k$-truss, $\alpha_{l}$ = LRC, and $\alpha_{tp}$ = triangle participation. The centrality measures are $\beta_d$ = Degree, $\beta_l$ = Local, $\beta_b$ = Betweenness, $\beta_c$ = Current-flow Closeness, $\beta_k$ = Katz, and $\beta_p$ = PageRank. The correlation measure are Spearman ($\rho_s$), Pearson ($\rho_p$), and Kendall Tau ($\tau_b$) from left to right.}
\label{fig:corrHeatmapsMain}

\end{figure*}

\subsubsection{Correlation Analysis}

The correlation analysis is conducted using Pearson, Spearman, and Kendall Tau for the 24 combinations of the 6 centrality and the 4 hierarchy measures. Figure \ref{fig:corrHeatmapsMain} reports the results for 6 networks illustrating the typical behavior of the 28 networks under investigation. The heatmap values range from the minimum correlation value (-0.007)  observed in the entire dataset to 1. The color spectrum ranges from dark blue to yellow. In the following, three categories in terms of correlation range are considered. Low correlation in the range -0.007 to 0.4 is associated to blue colors in the heatmaps. Medium correlation values ranging from 0.4 to 0.8 are represented by green colors. Finally, high correlation values above 0.8 are colored in yellow. 
Let's look at Spearman correlation results presented in the left hand side of figure \ref{fig:corrHeatmapsMain}. The heatmaps of 6 typical networks illustrating the results of this experiment are arranged from overall low correlation to overall high correlation.

The first typical behavior is illustrated  by the heatmap of the CS Ph.D. collaboration network. It is also observed in the EuroRoads network. In this case a large majority of correlation values are in the low and medium range. This translates in a heatmap where blue and dark green colors dominate. For these networks, we can conclude that there is no correlation between hierarchy and centrality measures. 
The second typical case is illustrated by the heatmap of E. Coli Transcription network. It concerns also all the biological networks (Yeast Protein, Human protein, Mouse Visual Cortex), the U.S. Power Grids and the Retweets Copenhagen networks. In this case, LRC and $k$-core are more or less well-correlated with the centrality measures. This translates in the heatmap with dominant colors ranging from light green to yellow. By contrast, low correlation values prevail for $k$-truss and triangle participation hierarchy measures. That is why dark green predominates in the heatmap.

The third case is illustrated by the heatmap of the Insects network. Mammals exhibit similar behavior. In this case, LRC and triangle participation are highly correlated with all the centrality measures. The dominant color of the heatmap is yellow for both hierarchy measures. $k$-core or $k$-truss show a quite different behavior with correlation values in the lower middle range. Indeed, dark green predominates in the heatmap. 

The fourth case is shown in the heatmap of the Physicians network. It is also observed in the U.S. States and Adolescent Health networks. In this case, correlation values range between 0.53 to 0.94. Light green and yellow are the predominant colors of the heatmap. $k$-truss is the less correlated hierarchy measure, with its dark green color. It is followed by triangle participation and $k$-core with their light green colors. Finally, LRC, mostly yellow on the heatmap appears to be well-correlated with other centrality measures.

The fifth case is presented using the heatmap of the Birds network. GrQc and Reptiles networks share the same behavior.  In this case, all the hierarchy measures exhibit high correlation values with the centrality measures. The dominant colors of the  heatmap are light green and yellow. However, betweenness centrality does not correlate well with the hierarchy measures. Indeed, the predominant color for the betweenness line of the heatmap is dark green with correlation values ranging from 0.35 to 0.54. 

The sixth and final case is represented by the heatmap of the World Metal Trade network. Zachary Karate Club, and Adjective Noun networks have quite similar heatmaps. In this case, almost all hierarchy and centrality measures are highly correlated. Dominant colors of the heatmaps are light green and yellow.

Turning to the heatmaps of Pearson and Kendall Tau's correlation, the results are presented respectively in the middle and the right hand columns of figure \ref{fig:corrHeatmapsMain}. Globally, it appears that Pearson correlation is closer to Spearman correlation values as compared to Kendall Tau. However, similar behavior is observed for both measures even if the correlation magnitudes are different. Note that the correlation magnitude of the Kendall Tau measure is systematically lower than Pearson and Spearman correlation corresponding values. This is not the case if Pearson is compared to Spearman. Indeed, Pearson correlation values can be higher or smaller than the corresponding Spearman correlation values. Consider for example the Insects network and the correlation between the degree centrality ($\beta_d$) and  $k$-core ($\alpha_c$) hierarchy measure. Spearman correlation is equal to 0.50. It is equal to 0.51 for  Pearson and it decreases 0.42 for Kendall Tau. 

Summing up the correlation analysis, one can say say that 6 different correlation trends are observed between hierarchy and centrality measures across the 28 networks. Furthermore, the three correlation measures lead to quite consistent results.

\begin{figure*}[ht!]
\begin{center}
\includegraphics[width=1\linewidth, height=2.6 in]{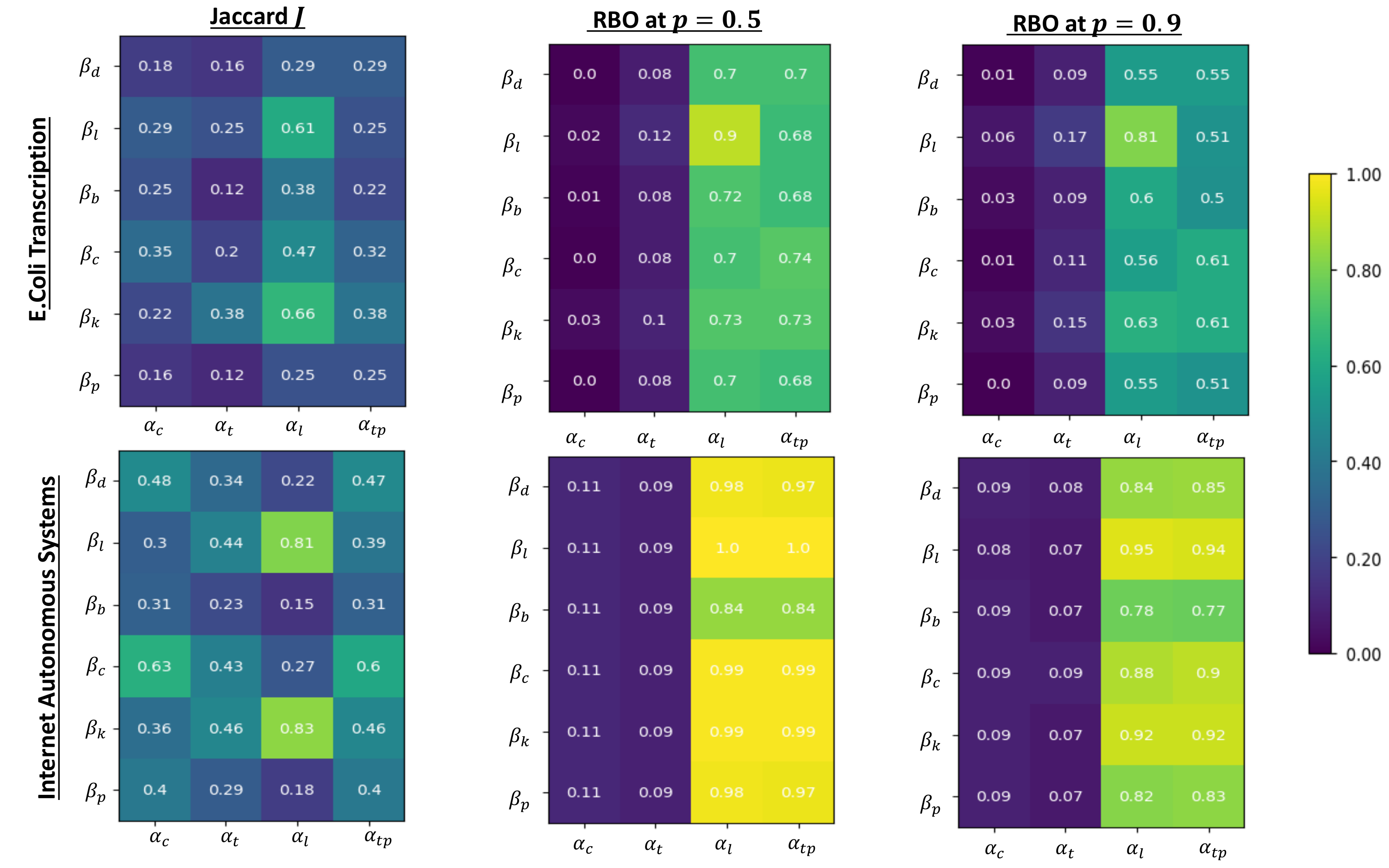}
\end{center}

\end{figure*}

\begin{figure*}[ht!]
\begin{center}
\includegraphics[width=1\linewidth, height=2.6 in]{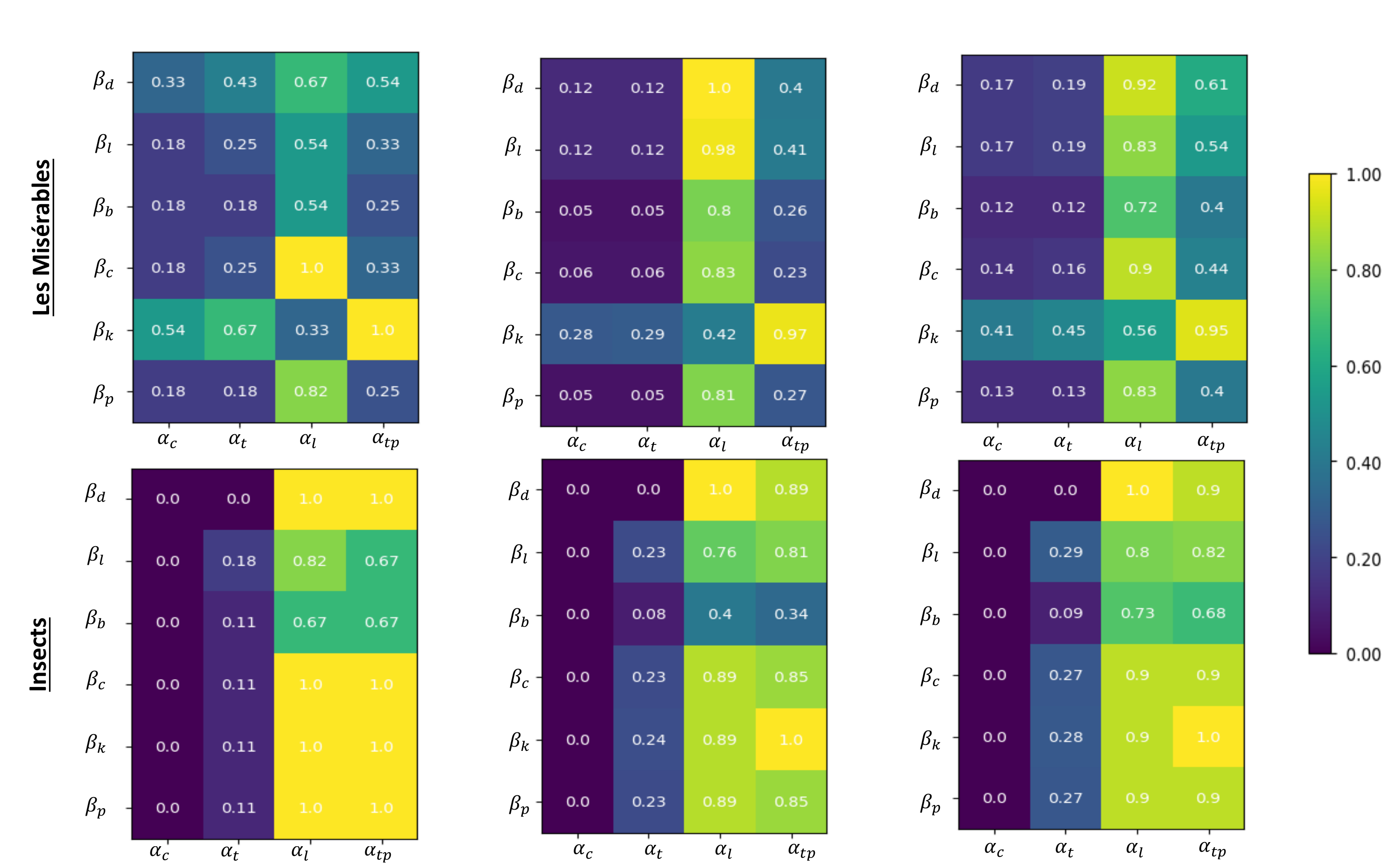}
\end{center}
\end{figure*}

\begin{figure*}[ht!]
\begin{center}
\includegraphics[width=1\linewidth, height=2.6 in]{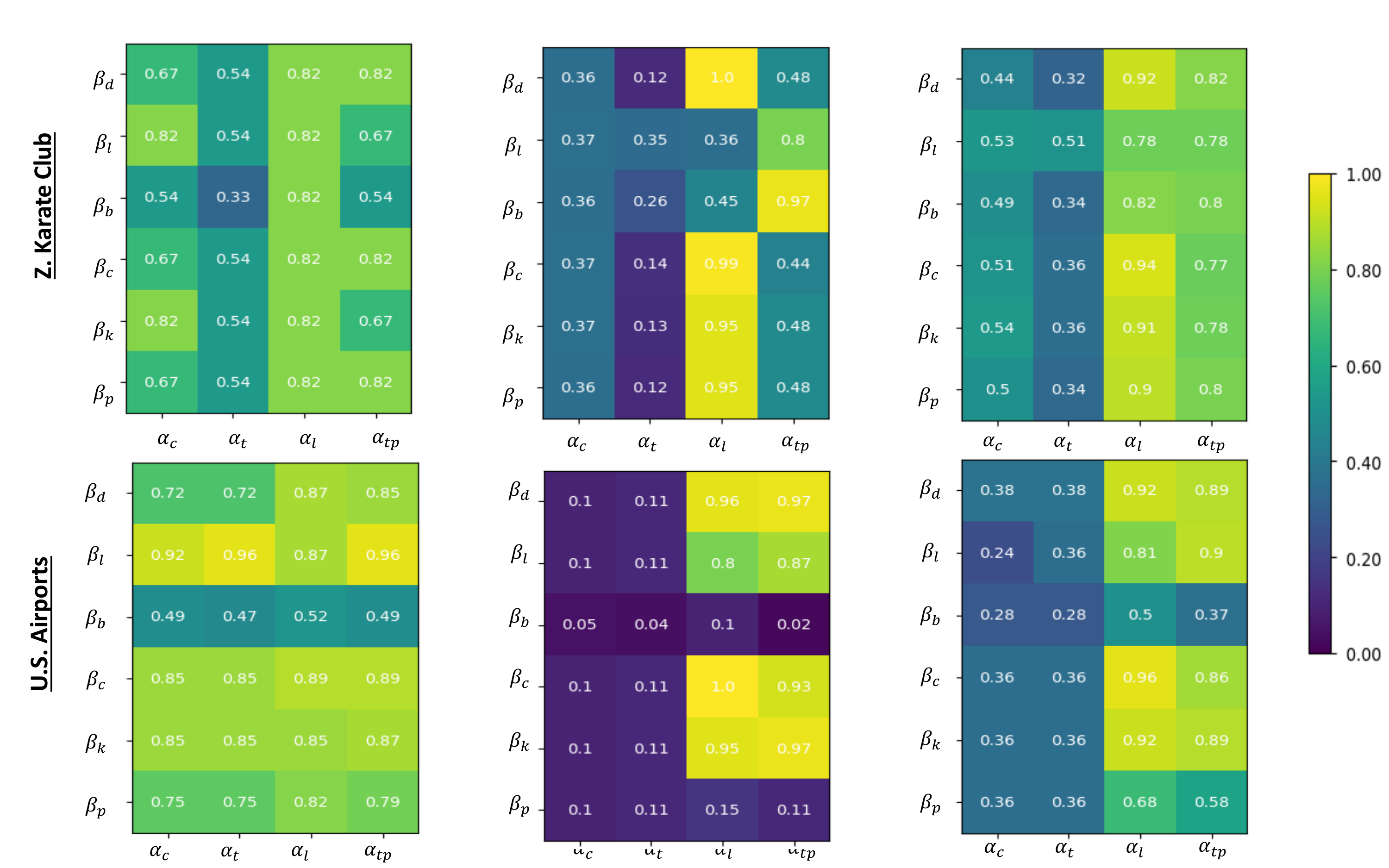}
\end{center}
\caption{Heatmaps of the similarity evaluation measures for the various combinations of hierarchy $\alpha_i$ and centrality $\beta_j$  measures of six real-world networks (from top to bottom). The hierarchy measures are $\alpha_{c}$ = $k$-core, $\alpha_{t}$ = $k$-truss, $\alpha_{l}$ = LRC, and $\alpha_{tp}$ = triangle participation. The centrality measures are $\beta_d$ = Degree, $\beta_l$ = Local, $\beta_b$ = Betweenness, $\beta_c$ = Current-flow Closeness, $\beta_k$ = Katz, and $\beta_p$ = PageRank. The similarity measures are Jaccard top-k nodes and RBO top-k nodes with $p=0.5$ and  $p=0.9$ from left to right.}
\label{fig:similarityHeatmapsMain}
\end{figure*}

\subsubsection{Similarity Analysis}
Results of similarity measures between the various hierarchy and centrality measures using the Jaccard index and RBO with its 4 versions have been computed. For the Jaccard index top 10\% nodes are used if the size of the network is $N \geq$ 150 nodes while top 10 nodes are used if  $N <$ 150. RBO is computed on the same datasets for comparative purpose. Two values of the parameter $p$ are used in RBO. A high value to give more weight to the top nodes ($p$=0.5) and  a medium value accounting for equal importance of the nodes ($p$=0.5). Furthermore RBO is computed using the top-k nodes and also the entire sample set. 

Figure \ref{fig:similarityHeatmapsMain} reports the results for the 6 networks (from top to bottom) representative of the various behavior observed in the experiments  and 3 out of 5 similarity measures (Jaccard, RBO top-k nodes with $p=0.5$ and RBO top-k nodes with $p=0.9$). The colors of the heatmaps vary from dark blue to yellow with similarity increasing from 0 to 1. The color range can be divided into 3 intervals. Blue indicates low similarity (0 to 0.4), green indicates medium similarity (0.4 to 0.8), and yellow indicates high similarity (0.8 to 1). Jaccard similarity is reported in the left-hand side of figure \ref{fig:similarityHeatmapsMain} to illustrate the various situations observed when evaluating the similarity of the 4 hierarchy measures with the 6 centrality measures. Networks are also arranged in increasing order between the two extreme cases (low similarity and high similarity).

E.Coli Transcription is the typical example chosen to illustrate the first category. Reptiles, Yeast, EuroRoads, U.S. Power Grids, NetSci, GrQc, and AstroPh exhibit a quite similar behavior. In this case, low to medium similarity is observed among almost all hierarchy and centrality combinations. In the heatmap blue and dark green predominates.

The second category is represented by the Internet Autonomous Systems network heatmap. Mouse Visual Cortex, U.S. States and Retweets Copenhagen networks belong also to this category. In this case, similarity of LRC with centrality measures is high. It ranges from low to medium for the other hierarchy measures ($k$-core, $k$-truss, triangle participation). Except for LRC, blue and dark green predominate in the heatmap.

The third category is illustrated by the heatmap of Les Mis\'erables. The heatmaps of Adjective Noun, PGP, and Facebook Politician Pages networks are quite comparable. In this case, the hierarchy measures show low similarity with a majority of centrality measures. However LRC and triangle participation can exhibit high similarity with the few centrality measures. Blue and green predominate in the heatmap with few yellow patches. 

The fourth category is represented by the heatmap of the Insects network. Mammals and Physicians networks are the other networks belonging to this category. In this case, $k$-core and $k$-truss have low similarity with all the centrality measures while LRC and triangle participation have high similarity. Blue and yellow predominate in the heatmaps.

The fifth category is illustrated by the Zachary Karate Club heatmap. It includes also Birds, World Metal Trade, Human Protein, and Facebook Ego networks. In this case most combinations between a centrality and a hierarchy measures exhibit similarity in a medium range. Green predominates in the heatmaps.

The sixth category contains a single network, the U.S. Airports network. It shows significant similarity across almost all possible combinations between hierarchy and centrality measures. Yet, betweenness centrality departs from this behavior. In this heatmap light green to yellow predominates except for the betweenness line.

 Results of the similarity evaluation using RBO top-k with $p$=0.5 and $p$=0.9 are reported in the middle and right-hand side of figure \ref{fig:similarityHeatmapsMain} respectively. As $p$ increases, more weight is given to the top overlapping nodes and their ranks. The same networks illustrating the 6 categories uncovered using the Jaccard similarity measure are used for comparative purposes. Remember that Jaccard doesn't take into account the rank and ties to measure the similarity between two sets, while RBO does. Indeed, a high RBO value means that the two sets share a high proportion of common nodes with similar ranking. Globally results are less mixed as compared to Jaccard similarity measure. 
 The six categories observed while using Jaccard similarity measure reduce to 3 categories using RBO top-k with $p$=0.5 (the middle column of figure \ref{fig:similarityHeatmapsMain}). 
 
 The first category is characterized by a very low similarity of $k$-core and $k$-truss with all the centrality measures. In contrast, LRC and triangle participation exhibit a medium to high similarity with all the centrality measures except some rare exceptions. This category regroups a high number of the networks under study. It is illustrated in figure \ref{fig:similarityHeatmapsMain} by the heatmaps of E.Coli Transcription, Internet Autonomous Systems, Insects, and U.S. Airports networks. Other networks that belong to this category are U.S. Power Grids, GrQc, Astroph, U.S. States, Retweets Copenhagen, Physicians, Birds, PGP, World Metal Trade, Madrid Train Bombings, and Mammals. Dark green predominates in the first two columns of the heatmap ($k$-core and $k$-truss), while light green and yellow predominate in the third and fourth column (LRC and triangle participation).

In the second category, LRC shows a high similarity with the centrality measures, while similarity with almost all centrality measures is low for $k$-core, $k$-truss, and triangle participation. Les Mis\'erables, reported in figure \ref{fig:similarityHeatmapsMain}, is a typical example of this behavior. Facebook Ego, Human Protein, and Mouse Visual Cortex are the other networks that belong to this category. Dark blue predominates in the heatmaps except on the third column (LRC).

\begin{figure}[ht!]
\begin{center}
\includegraphics[width=0.5\linewidth, height= 1.5 in]{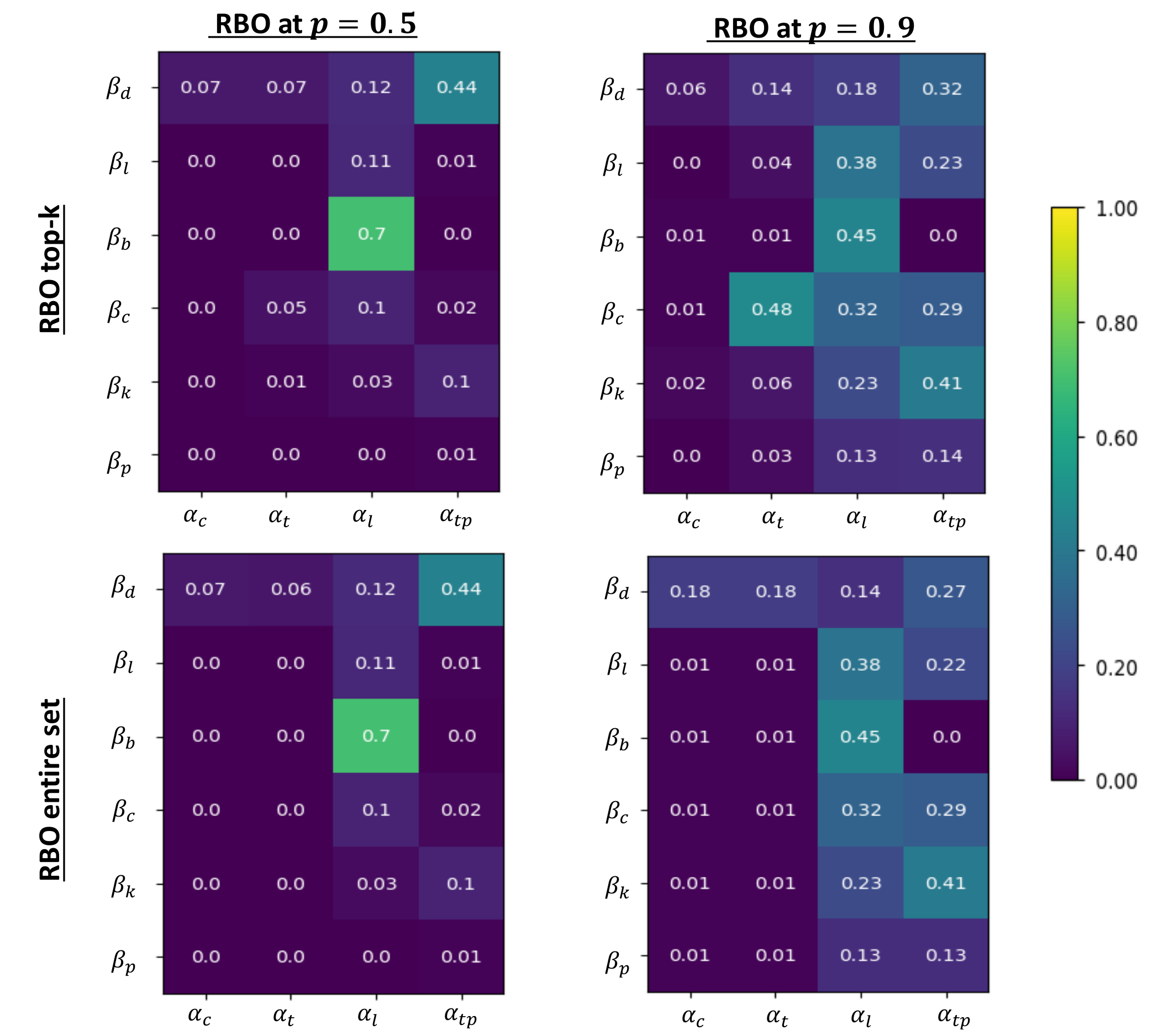}
\end{center}
\caption{Heatmaps of RBO top-k and RBO of entire set (top to bottom) with $p=0.5$ and  $p=0.9$ (left to right) for the various combinations of hierarchy $\alpha_i$ and centrality $\beta_j$  measures of the Adolescent Health Network. The hierarchy measures are $\alpha_{c}$ = $k$-core, $\alpha_{t}$ = $k$-truss, $\alpha_{l}$ = LRC, and $\alpha_{tp}$ = triangle participation. The centrality measures are $\beta_d$ = Degree, $\beta_l$ = Local, $\beta_b$ = Betweenness, $\beta_c$ = Current-flow Closeness, $\beta_k$ = Katz, and $\beta_p$ = PageRank.}
\label{fig:AdolescentHealthRBOTopKVSALLAndATrendByItself}
\end{figure}

In the third category one can observe low to medium similarity among all hierarchy and centrality measures. It is illustrated in figure \ref{fig:AdolescentHealthRBOTopKVSALLAndATrendByItself} by the heatmap of the Adolescent Health network. Other networks that fall into this category are Reptiles, Yeast Protein, EuroRoads, NetSci, CS Ph.D., and Facebook Politician Pages. Dark blue predominates the whole heatmap.

Let's now compare RBO top-k at $p$=0.5 to RBO top-k at $p$=0.9. Globally the results are quite comparable. However, a closer look allows to distinguish three cases for a given hierarchy and centrality combination. 

In the first case, increasing the $p$ value increases the similarity value. Indeed, top nodes are given more importance and as they are identical and with the same rank, hence the similarity of the two sets increases. The combination of LRC hierarchy and PageRank centrality of the U.S. Airports network shown in figure \ref{fig:similarityHeatmapsMain} is a typical example. One can notice that $RBO_{p=0.5}(\alpha_l, \beta_p)$=0.15 increases to $RBO_{p=0.9}(\alpha_l, \beta_p)$=0.68. 

In the second case, the similarity value decreases if $p$ increases. Therefore, there is less top nodes identical in the two sets or their ranking differs. This results in a lower similarity for high $p$ value. The combination between triangle participation hierarchy and degree centrality for the E.Coli Transcription network shown in figure \ref{fig:similarityHeatmapsMain} illustrates this situation. Indeed, RBO decreases as $p$ increases ($RBO_{p=0.5}(\alpha_{tp}, \beta_d)$=0.7,  $RBO_{p=0.9}(\alpha_{tp}, \beta_d)$=0.55). 

In the third case, similarity values at $p$=0.5 and at $p$=0.9 are identical. In this case, the overlap of nodes and their rankings do not significantly differ. Hence, increasing $p$ does not have a notable effect on the similarity value. Such cases are observed for extreme values. An example of this case is the combination of $k$-core hierarchy and degree centrality in the Insects network in figure \ref{fig:similarityHeatmapsMain}, where $RBO_{p=0.5}(\alpha_{c}, \beta_d)$=0 whereas $RBO_{p=0.9}(\alpha_{c}, \beta_d)$=0. Another extreme case in the same network concerns the similarity between LRC hierarchy and degree centrality, where $RBO_{p=0.5}(\alpha_{l}, \beta_d)$=1 whereas $RBO_{p=0.9}(\alpha_{l}, \beta_d)$=1. Note that the numbers are rounded so there may be a very small difference in the values. Yet, this is the characteristic of the third case, where we almost have the same similarity value regardless of the value of $p$. 
 
Finally, lets compare RBO top-k nodes to RBO on the entire set of nodes. The results are quite similar. In other words, comparing the top-k nodes or all the nodes in the network does not provide much more information regarding the similarity of the hierarchy and centrality combinations. Figure \ref{fig:AdolescentHealthRBOTopKVSALLAndATrendByItself} illustrates this observation with the Adolescent Health network. The complementary results of  RBO using the entire dataset using the other networks are provided in supplementary materials.

Summing up the similarity analysis, using the  Jaccard similarity one can classify  the 28 networks used into  6 categories. This reduces to 3 categories when RBO is used.  The first set of experiments do not allow to get a full picture of the relations between centrality and hierarchy measures. First of all, there is no clear relationship between correlation and similarity measures. One can notice that variations can be observed across networks. Nevertheless, some general trends seems to be emerging. Differences between the various centrality measures are not clear-cut, except for betweenness that sometimes behave quite differently than the others. Another remark is that $k$-core and $k$-truss exhibit more often low similarity as compared to LRC and triangle participation. Finally, it appears that for all the correlation measures and the Jaccard similarity the results are more mixed as compared to RBO.

\subsection{Comparing the networks according to the evaluation measures sample sets}

In order to relate the interactions between centrality and hierarchy measures to the network structure, we conduct three experiments based on the results of the previous one. In the first experiment, given an evaluation measure (correlation or similarity), the Pearson correlation is computed between the sample sets of the various combinations between centrality and hierarchy. The second experiment ranks the networks and the third one categorizes the networks based on the $k$-means algorithm. 

\subsubsection{Correlation Analysis of the evaluation measures sample sets}
\begin{figure*}[h]
\begin{center}
\includegraphics[width=1\linewidth, height=5.5 in]{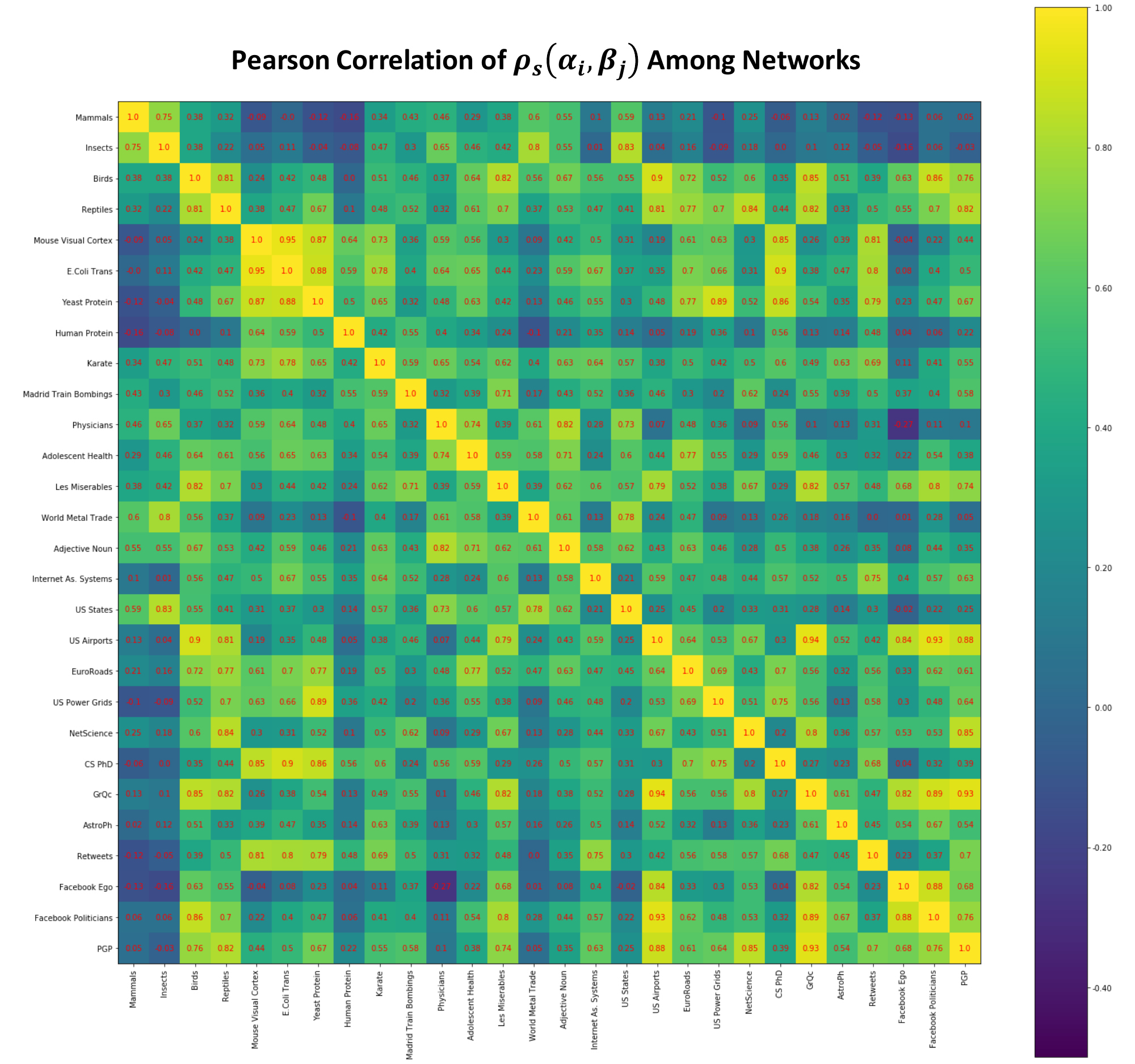}
\end{center}
\caption{Heatmap of the Pearson correlation between the network evaluation measures sample sets $\rho_p(\Gamma_i , \Gamma_j)$ for all the pairs of the 28 networks $S_i, S_j$. In the sample set $\Gamma_i= \{\gamma(\alpha_1,\beta_1), \gamma(\alpha_1,\beta_2),..., \gamma(\alpha_i,\beta_j)\}$, the hierarchy measures are $\alpha_{c}$ = $k$-core, $\alpha_{t}$ = $k$-truss, $\alpha_{l}$ = LRC, and $\alpha_{tp}$ = triangle participation. The centrality measures are $\beta_d$ = Degree, $\beta_l$ = Local, $\beta_b$ = Betweenness, $\beta_c$ = Current-flow Closeness, $\beta_k$ = Katz, and $\beta_p$ = PageRank. The evaluation measure $\Gamma$ is the Spearman correlation measure.}
\label{fig:PearsonCorrOfNetworksAccordingToSpearmanCorr}
\end{figure*}

For a given evaluation measure $\gamma$, a network $S_i$ is characterized by the sample set of the evaluation measures' values between all the centrality and hierarchy measures $\Gamma_i= \{\gamma(\alpha_1,\beta_1), \gamma(\alpha_1,\beta_2),..., \gamma(\alpha_i,\beta_j)\}$. In order to compare the networks two-by-two, the Pearson correlation between the sample sets $\rho_p(\Gamma_i , \Gamma_j)$ is computed. Experiments are performed using the three correlation measures and the two similarity measures. Results are reported for the Spearman correlation used as the evaluation measure in figure \ref{fig:PearsonCorrOfNetworksAccordingToSpearmanCorr}.
 One can observe that the heatmap is very patchy. It does not contain any clear pattern. Few yellow spots emerge now and then in a high proportion of green and blue. In other words, the vast majority of correlations are in the low and medium range. Nevertheless, some networks show strong correlation. Results using the alternative correlation evaluation measures (Pearson and Kendall Tau) are reported in supplementary materials. Indeed, they are in the same vein and do not convey much more information.
 
\begin{figure*}[h]
\begin{center}
\includegraphics[width=1\linewidth, height=5.5 in]{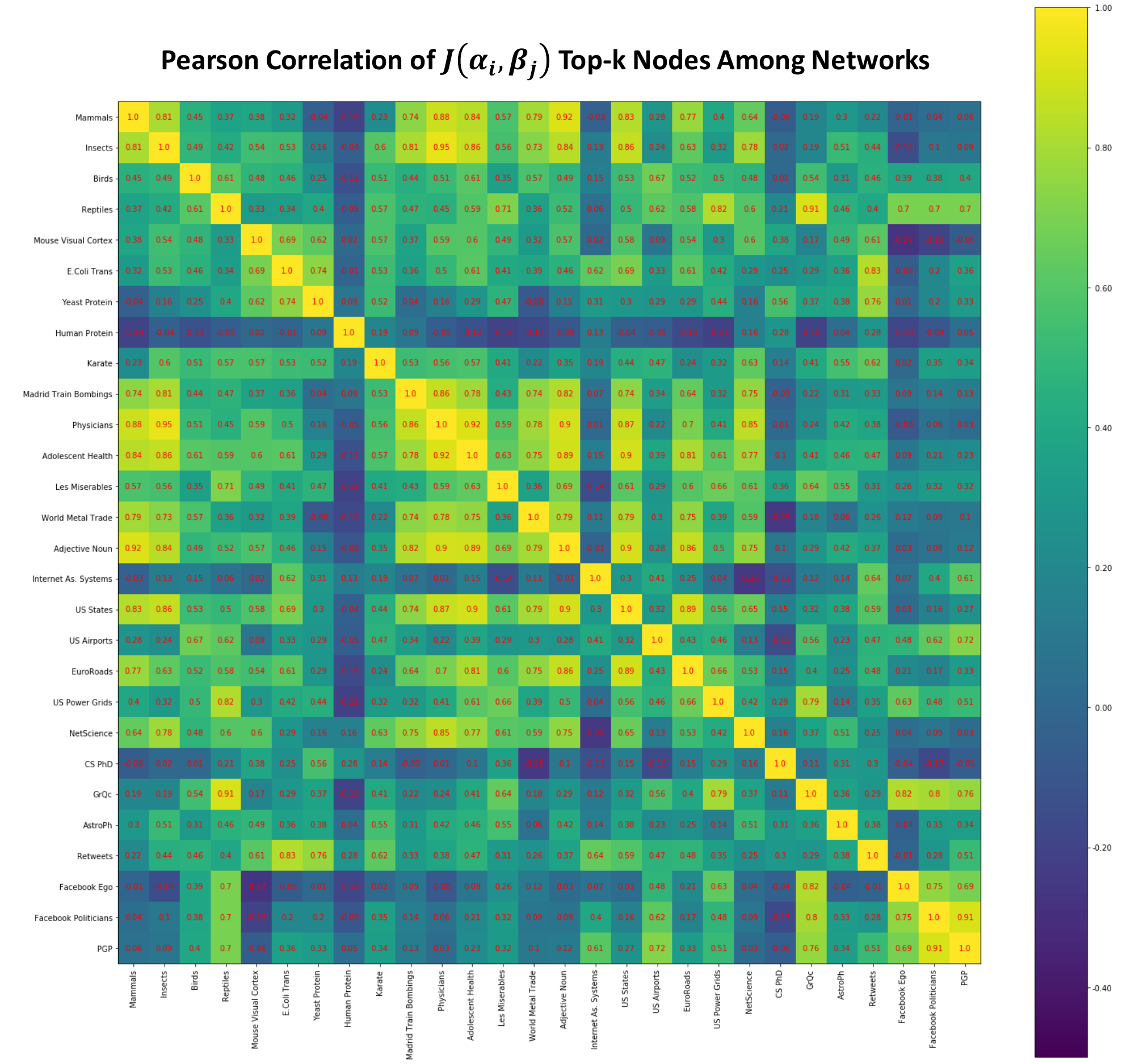}
\end{center}
\caption{Heatmap of the Pearson correlation between the network evaluation measures sample sets $\rho_p(\Gamma_i , \Gamma_j)$ for all the pairs of the 28 networks $S_i, S_j$. In the sample set $\Gamma_i= \{\gamma(\alpha_1,\beta_1), \gamma(\alpha_1,\beta_2),..., \gamma(\alpha_i,\beta_j)\}$, the hierarchy measures are $\alpha_{c}$ = $k$-core, $\alpha_{t}$ = $k$-truss, $\alpha_{l}$ = LRC, and $\alpha_{tp}$ = triangle participation. The centrality measures are $\beta_d$ = Degree, $\beta_l$ = Local, $\beta_b$ = Betweenness, $\beta_c$ = Current-flow Closeness, $\beta_k$ = Katz, and $\beta_p$ = PageRank. The evaluation measure $\Gamma$ is the Jaccard similarity measure.}
\label{fig:PearsonCorrOfNetworksAccordingToJaccardSim}
\end{figure*}

Let's now turn to the similarity measures. Figure \ref{fig:PearsonCorrOfNetworksAccordingToJaccardSim} reports the results for the Jaccard evaluation measure. Globally, the results does not depart from the previous ones. However, the values of the correlation are much smaller and some patterns appear more clearly. For example, Facebook Ego, CS Ph.D. and Human Protein networks appear quite uncorrelated with a vast majority of the other networks.

\begin{figure*}[h]
\begin{center}
\includegraphics[width=1\linewidth, height=5.5 in]{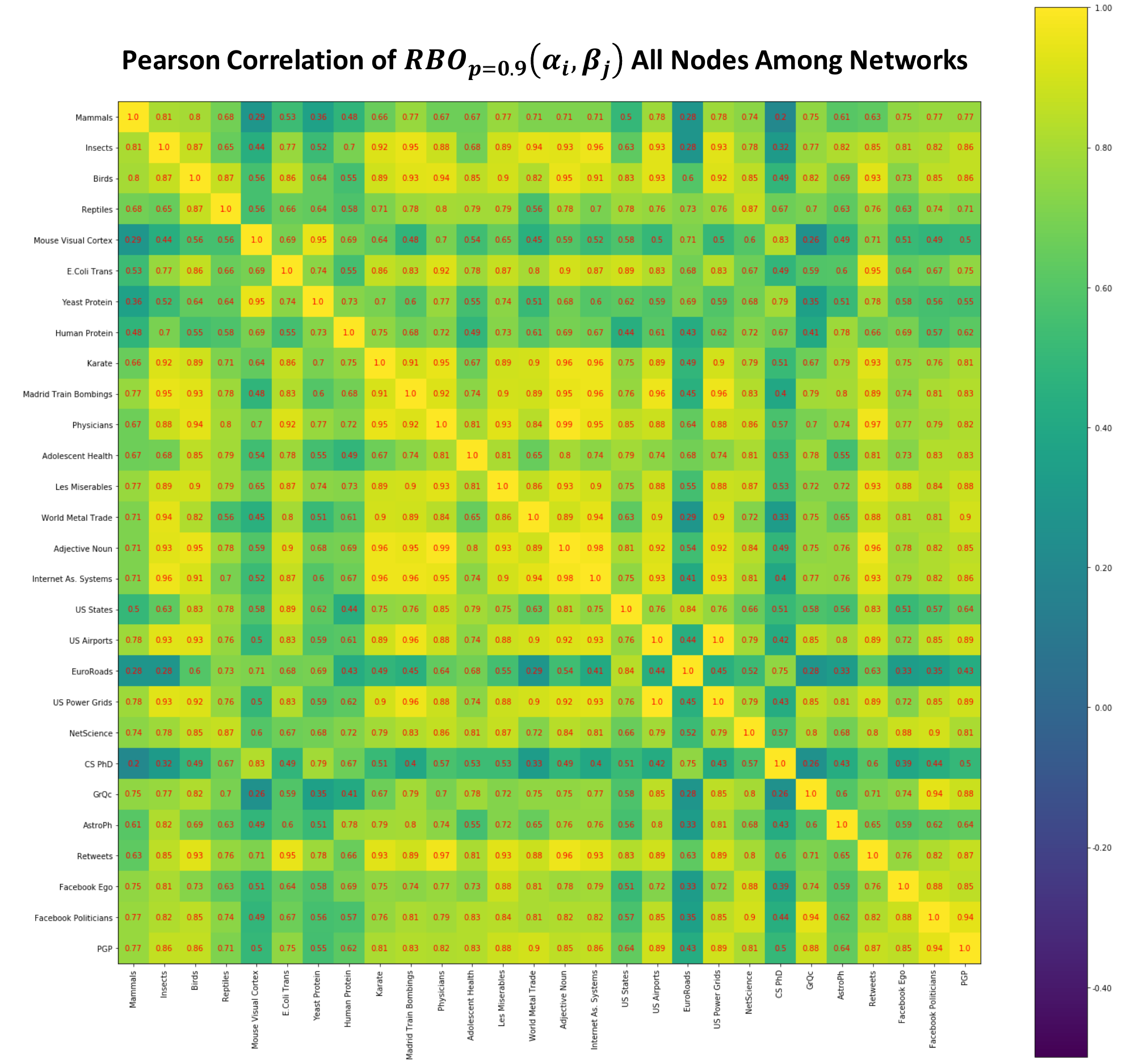}
\end{center}

\caption{Heatmap of the Pearson correlation between the network evaluation measures sample sets $\rho_p(\Gamma_i , \Gamma_j)$ for all the pairs of the 28 networks $S_i, S_j$. In the sample set $\Gamma_i= \{\gamma(\alpha_1,\beta_1), \gamma(\alpha_1,\beta_2),..., \gamma(\alpha_i,\beta_j)\}$, the hierarchy measures are $\alpha_{c}$ = $k$-core, $\alpha_{t}$ = $k$-truss, $\alpha_{l}$ = LRC, and $\alpha_{tp}$ = triangle participation. The centrality measures are $\beta_d$ = Degree, $\beta_l$ = Local, $\beta_b$ = Betweenness, $\beta_c$ = Current-flow Closeness, $\beta_k$ = Katz, and $\beta_p$ = PageRank. The evaluation measure $\Gamma$ is the RBO similarity measure of the entire set of nodes with $p$=0.9.}
\label{fig:PearsonCorrofRBOAllSimOfNetworksAtP09}
\end{figure*}

A less patchy matrix is observed using RBO as an evaluation measure. Especially when considering a high value of $p$ ($p$=0.9) as shown in figure \ref{fig:PearsonCorrofRBOAllSimOfNetworksAtP09}. This confirms that the results are more consistent when RBO is used as an evaluation measure. The results for the other configurations of RBO are quite convergent. They are given in supplementary materials.

\subsubsection{Grouping the networks based on the binarized evaluation measures sample sets}

In order to get a clearer picture about the relations between the network topology and the evaluation measures, the values of the evaluation measures $\gamma(\alpha_i,\beta_j)$ are binarized for each network. In other words, rather than using the continuous values of the evaluation measures, two cases are considered. Based on a threshold ($\mu$=0.7), the evaluation measure for a given combination of centrality and hierarchy is set to 1 (meaningful) or 0 (not meaningful). Then the proportion of meaningful values out of the 24 combinations between centrality and hierarchy is computed for each network. Networks are then ranked according to the proportion of meaningful values, and topological properties of the networks are explored in order to check if an order relation is also uncovered. Finally, based on these relations the networks are categorized. This set of experiments is performed for both the correlation and the similarity measures.

Table \ref{table:new2} shows the networks ranked according to the meaningful values of the three correlation measures together with their basic topological properties $($density ($\nu$),  transitivity ($\zeta$) and assortativity ($k_{nn}(k)$)$)$. The rankings based on each of the correlation measures (Spearman, Pearson, and Kendall Tau) are provided in the supplementary materials. Overall those values are well correlated (see table \ref{table:corrofrankings}) and consequently the ranking of the networks are quite similar except for few exceptions.

Looking at this table, we can clearly divide the networks into 3 groups based on their topological characteristics. The first group is made of the 8 networks with the higher ranks (From Adjective Noun to Mammals). One can notice that these networks exhibit in general high density ($\nu$), high transitivity ($\zeta$), and negative or near-zero assortativity ($k_{nn}(k)$).The second group contains 10 networks ranked in the middle range (Physicians to PGP). It is characterized in general by low density ($\nu$), high transitivity ($\zeta$), and positive assortativity ($k_{nn}(k)$). Finally, the third group is made of the 10 networks with the lowest ranks. Its typical features are low density ($\nu$), low transitivity ($\zeta$), and negative or near-zero assortativity ($k_{nn}(k)$).

The process for relating the topological properties of the networks to their ranking according to the proportion of meaningful similarity values is slightly different than what has been done with the correlation evaluation measures. Table \ref{table:corrofrankings} reports the correlation between the various version of  RBO and Jaccard similarity measure. One can see that correlation is very low. Indeed, rankings according to Jaccard similarity are quite different than rankings based on RBO. This is the reason why we prefer to consider Jaccard and RBO separately. Consequently, we do not aggregate the ranking of Jaccard and RBO to get an overall rank for each network.

\begin{table}[ht!]
  \centering
  \caption{Real-world networks sorted in decreasing order according to the proportion of meaningful correlation measures ($|\lambda_{\rho_s}$ +$\lambda_{\rho_p}$ +$\lambda_{\tau_b}| \geq$ 0.7) denoted as $\lambda_c$. The network basic topological properties are reported ($\nu$ is the density, $\zeta$ is the transitivity and $k_{nn}(k)$ is the assortativity).}
   \label{table:new2}    
      \begin{tabular}{p{4cm}cccc}
      \toprule
    Network &  $\lambda_c$ &   $\nu$  & $\zeta$ & $k_{nn}(k)$ \\
    \midrule
    Adjective Noun & 59$/$72 & 0.068 & 0.156 & -0.129
    \\
      Zachary Karate & 55$/$72  & 0.139 & 0.255 & -0.475
     \\
      Les Mis\'erables  & 55$/$72  & 0.086 &  0.498 & -0.165
     \\
    World Metal Trade & 54$/$72 & 0.276 & 0.459 & -0.391
     \\
    U.S. Airports & 52$/$72 & 0.023 & 0.351 & -0.267
     \\
  Madrid Train Bomb. &  51$/$72 & 0.120 & 0.561 & 0.029
     \\
   Birds & 51$/$72 & 0.577 & 0.472 & 0.062
     \\
   Mammals & 49$/$72 & 0.716 & 0.727 & -0.004
     \\
  \hline 
   Physicians & 41$/$72  & 0.068 & 0.174 & -0.084
     \\
  Facebook Pol. Pages  &  39$/$72 & 0.002 & 0.301 & 0.018
     \\
   Facebook Ego & 38$/$72 & 0.010 & 0.519 & 0.063
     \\
    Insects &  36$/$72 & 0.798 & 0.785 & -0.030
     \\
  U.S. States & 34$/$72 & 0.090 & 0.406 & 0.233
     \\
  AstroPh & 34$/$72  & 0.001 & 0.317 & 0.201
     \\
   GrQc    & 33$/$72 & 0.001 & 0.628 & 0.639
     \\
  Adolescent Health  & 29$/$72 &  0.002 & 0.141 & 0.231
     \\
   Reptiles   & 26$/$72 & 0.008 & 0.419 & 0.342
     \\
   PGP & 25$/$72 & 0.0004 & 0.378 & 0.238
     \\
\hline 
    Retweets Copenhagen & 23$/$72 & 0.003 & 0.060 & -0.099
     \\
   Internet A. Systems & 20$/$72 & 0.0006 & 0.009 & -0.181
     \\
  NetSci   & 19$/$72 & 0.012 & 0.430 & -0.081
     \\
 Human Protein & 19$/$72 &  0.002 & 0.007 & -0.331
     \\
  E.Coli Transcription  & 19$/$72 & 0.008 & 0.023 & -0.263
     \\
  Mouse Vis. Cortex  & 13$/$72 & 0.011 & 0.004 & -0.844
   \\
 Yeast Protein  & 12$/$72 & 0.001 & 0.051 & -0.207
 \\
    U.S. Power Grids   & 5$/$72 & 0.0005 & 0.103 & 0.003
     \\
     EuroRoads   & 4$/$72  & 0.002 & 0.035 & 0.090
     \\
    CS Ph.D.  & 3$/$72  & 0.001 & 0.002 & -0.253
     \\

     \bottomrule
      \end{tabular}
\end{table}

\begin{table}[ht!]
   \centering
    \caption{Pearson correlation between the rankings of the networks based on their meaningful evaluation measures.Correlation measures are Spearman ($|\rho_s(\alpha, \beta)| \geq$ 0.7) denoted as $\lambda_{\rho_s}$, Pearson ($|\rho_p(\alpha, \beta)| \geq$ 0.7) denoted as $\lambda_{\rho_p}$, Kendall Tau ($|\tau_b(\alpha, \beta)| \geq$ 0.7) denoted as $\lambda_{\tau_b}$.  Similarity measures are Jaccard ($|J(\alpha, \beta)| \geq$ 0.7) denoted as $\lambda_J$, RBO at $p$=0.5 ($|RBO_{p=0.5}(\alpha, \beta)| \geq$ 0.7) denoted as $\lambda^{_{p=0.5}}_{RBO}$, and finally RBO at $p$=0.9 ($|RBO_{p=0.9}(\alpha, \beta)| \geq$ 0.7) denoted as $\lambda^{_{p=0.9}}_{RBO}$.}
    \label{table:corrofrankings}   
      \begin{tabular}{ccccccc}
      \toprule

      & $\lambda_{\rho_s}$  &  $\lambda_{\rho_p}$  & $\lambda_{\tau_b}$  & $\lambda_J$ & $\lambda^{_{p=0.5}}_{RBO}$  & $\lambda^{_{p=0.9}}_{RBO}$\\
    \midrule
    $\lambda_{\rho_s}$ & 1 
    \\
    $\lambda_{\rho_p}$ & 0.84  & 1
     \\
     $\lambda_{\tau_b}$ & 0.90 & 0.88 & 1 
     \\
    $\lambda_{J}$ & 0.78 & 0.67 & 0.73 & 1
     \\
    $\lambda^{_{p=0.5}}_{RBO}$ & 0.28 & 0.08 & 0.25 & 0.32 & 1 
     \\
    $\lambda^{_{p=0.9}}_{RBO}$ & 0.45 & 0.32 & 0.47 & 0.50 & 0.78 & 1
     \\
 
     \bottomrule
      \end{tabular}
\end{table}

Results for the ranked networks according to the proportion of meaningful Jaccard values together with their respective topological characteristics are given in table \ref{table:rankingSigSimJaccard}. One can see that the networks can be divided into 2 groups based on their topological characteristics. The first group of 7 top ranked networks (from US Airports to World Metal Trade) is characterized by high density ($\nu$) and high transitivity ($\zeta$), with negative or near-zero assortativity ($k_{nn}(k)$). The second group is made of the 21 remaining networks. They exhibit low or even zero proportion of meaningful Jaccard similarity, and are characterized by low density ($\nu$). The other properties do not show clear trends. Indeed, transitivity ($\zeta$) and assortativity ($k_{nn}(k)$) fluctuate in a wide range.

For RBO, the proportion of meaningful similarities according to RBO top-k for the two $p$ values ($RBO_{p=0.5}$ and $RBO_{p=0.9}$) are averaged. indeed, both ranking are well-correlated. The detailed rankings of each evaluation measure considered individually are provided in the supplementary materials. The rankings of the networks with their topological properties are given in table \ref{table:new4}. In this case, the networks can also be divided into 2 groups based on their topological characteristics. The first group is composed of the 12 top ranked networks (from Adjective Noun to Mammals). These networks exhibit high transitivity ($\zeta$). Their density and transitivity values fluctuate in a wide range, with no particular visible trend. The 16 remaining networks forming the second category are characterized by low density ($\nu$). No particular pattern is observed in their transitivity and assortativity values which vary in a wide range. Note that the rankings obtained with RBO using the entire node set provides very similar rankings to RBO top-k. That is the reason why they are presented in supplementary materials.

 Table \ref{table:birdseyeviewofgroups} summarizes the common properties shared by the various groups uncovered using the various type of evaluation measures. The density of the networks seems to be the most discriminating feature allowing to categorize the networks. Indeed, if one relies only on this parameter two categories emerge (high density, low density). This behavior is well-pronounced for correlation and Jaccard evaluation measures. 
 The high density category made of the highly ranked networks is made of the same 6 networks out of 8 for correlation and 7 out of 8 for Jaccard. Comparing the 16 networks belonging to the low density category according to RBO to the corresponding category based on Jaccard and correlation evaluation measures, show that the results are also quite consistent. Fifteen out of sixteen belongs to the low density category uncovered by the correlation evaluation measure and all of them belong to the low density category uncovered by the Jaccard evaluation measure. Transitivity is also an important feature influencing the classification. One can notice that, whatever the evaluation measure used, the category made of the top ranked networks always group networks with high transitivity. Finally, the groupings are less sensitive to assortativity except for the correlation evaluation measures.
It is worth noticing that whatever the evaluation measure the networks belonging to the groups made of the top ranked networks exhibit high transitivity. Furthermore, the networks belonging to the groups made of the low ranked networks are sharing a low density. Hence, these two characteristics may be good predictors for two different groups. To sum up, these results allow us to conclude that the global topological characteristics of a network affect the relationship between hierarchy and centrality. Correlation is the most sensitive evaluation measures to highlight these relations, followed by Jaccard and RBO.

\begin{table}[ht!]
   \centering
    \caption{Real-world networks sorted in decreasing order according to the proportion of meaningful Jaccard similarity measures $|J(\alpha, \beta)| \geq$ 0.7 denoted as $\lambda_J$. The network basic topological properties are reported ($\nu$ is the density, $\zeta$ is the transitivity and $k_{nn}(k)$ is the assortativity).}
    \label{table:rankingSigSimJaccard}
      \begin{tabular}{p{4cm}cccc}
      \toprule

    Network &  $\lambda_J$  &   $\nu$  & $\zeta$ & $k_{nn}(k)$ \\
    \midrule
   U.S. Airports & 20$/$24 & 0.023 & 0.351 & -0.267
     \\
   Zachary Karate & 11$/$24   & 0.139 & 0.255 & -0.475
     \\
    Insects & 9$/$24 & 0.798 & 0.785 & -0.030
     \\
    Madrid Train Bomb. &  8$/$24  & 0.120 & 0.561 & 0.029
     \\
   Birds & 7$/$24 & 0.577 & 0.472 & 0.062
     \\
   Mammals & 5$/$24  & 0.716 & 0.727 & -0.004
     \\
    World Metal Trade & 5$/$24 & 0.276 & 0.459 & -0.391
     \\
\hline 
   Physicians & 5$/$24   & 0.068 & 0.174 & -0.084
     \\
   Facebook Ego & 4$/$24 & 0.010 & 0.519 & 0.063
     \\
   Mouse Vis. Cortex  & 3$/$24  & 0.011 & 0.004 & -0.844
   \\
   Les Mis\'erables  & 3$/$24  & 0.086 &  0.498 & -0.165
    \\
    Retweets Copenhagen & 3$/$24 & 0.003 & 0.060 & -0.099
     \\
   Internet A. Systems & 2$/$24  & 0.0006 & 0.009 & -0.181
     \\
   U.S. States & 2$/$24 & 0.090 & 0.406 & 0.233
     \\
   Human Protein & 1$/$24  &  0.002 & 0.007 & -0.331
     \\
    CS Ph.D.  & 1$/$24  & 0.001 & 0.002 & -0.253
     \\
   Facebook Pol. Pages  & 1$/$24  & 0.002 & 0.301 & 0.018
     \\
   PGP & 1$/$24  & 0.0004 & 0.378 & 0.238
     \\
   Reptiles   & 0$/$24 & 0.008 & 0.419 & 0.342
     \\
   E.Coli Transcription  & 0$/$24  & 0.008 & 0.023 & -0.263
     \\
   Yeast Protein  & 0$/$24 & 0.001 & 0.051 & -0.207
     \\
    Adolescent Health  & 0$/$24 & 0.002 & 0.141 & 0.231
     \\
    Adjective Noun  & 0$/$24  & 0.068 & 0.156 & -0.129
     \\
    EuroRoads   & 0$/$24 & 0.002 & 0.035 & 0.090
     \\
    U.S. Power Grids   & 0$/$24  & 0.0005 & 0.103 & 0.003
     \\
   NetSci   & 0$/$24 & 0.012 & 0.430 & -0.081
     \\
   GrQc    & 0$/$24  & 0.001 & 0.628 & 0.639
     \\
   AstroPh & 0$/$24  & 0.001 & 0.317 & 0.201
     \\
 
     \bottomrule
      \end{tabular}
\end{table}

\begin{table}[ht!]
   \centering
    \caption{Real-world networks sorted in decreasing order according to the proportion of meaningful RBO similarity ($|\lambda_{RBO_{p=0.5}}$ +$\lambda_{RBO_{p=0.9}}| \geq$ 0.7) denoted as $\lambda_{RBO}$. The network basic topological properties are reported ($\nu$ is the density, $\zeta$ is the transitivity and $k_{nn}(k)$ is the assortativity).}
    \label{table:new4}    
       \begin{tabular}{p{4cm}cccc}
      \toprule
    Network &  $\lambda_{RBO}$ &   $\nu$  & $\zeta$ & $k_{nn}(k)$ \\
    \midrule
     Adjective Noun  & 24$/$48 & 0.068 & 0.156 & -0.129
     \\  
   Internet A. Systems &  24$/$48 & 0.0006 & 0.009 & -0.181
     \\
    World Metal Trade & 21$/$48 & 0.276 & 0.459 & -0.391
     \\
    Insects & 21$/$48  & 0.798 & 0.785 & -0.030
     \\
    Madrid Train Bomb. &  20$/$48 & 0.120 & 0.561 & 0.029
     \\
  Physicians & 19$/$48  & 0.068 & 0.174 & -0.084
     \\
    Zachary Karate & 18$/$48 & 0.139 & 0.255 & -0.475
     \\
    U.S. Airports & 16$/$48 & 0.023 & 0.351 & -0.267
     \\
    U.S. Power Grids  & 16$/$48  & 0.0005 & 0.103 & 0.003
     \\
  Birds &  15$/$48   & 0.577 & 0.472 & 0.062
     \\
 AstroPh & 14$/$48 & 0.001 & 0.317 & 0.201
     \\
   Mammals & 13$/$48 & 0.716 & 0.727 & -0.004
     \\
\hline 
   Les Mis\'erables  & 12$/$48 & 0.086 &  0.498 & -0.165
    \\
   U.S. States & 12$/$48 & 0.090 & 0.406 & 0.233
     \\
Retweets Copenhagen & 12$/$48 & 0.003 & 0.060 & -0.099
     \\
   PGP & 12$/$48 & 0.0004 & 0.378 & 0.238
     \\
   E.Coli Transcription  & 10$/$48 & 0.008 & 0.023 & -0.263
     \\  
   GrQc    & 9$/$48 & 0.001 & 0.628 & 0.639
     \\
   Human Protein & 7$/$48 &  0.002 & 0.007 & -0.331
     \\
     NetSci   & 5$/$48 & 0.012 & 0.430 & -0.081
     \\  
    Facebook Ego & 9$/$48 & 0.010 & 0.519 & 0.063
     \\  
   Mouse Vis. Cortex  & 3$/$48 & 0.011 & 0.004 & -0.844
   \\
   Yeast Protein  & 3$/$48 & 0.001 & 0.051 & -0.207
     \\
     Reptiles   & 2$/$48 & 0.008 & 0.419 & 0.342
     \\
  CS Ph.D.  & 2$/$48 & 0.001 & 0.002 & -0.253
     \\
   Adolescent Health  & 1$/$48  & 0.002 & 0.141 & 0.231
     \\
  Facebook Pol. Pages  & 1$/$48 & 0.002 & 0.301 & 0.018
     \\
    EuroRoads   & 1$/$48 & 0.002 & 0.035 & 0.090
     \\

     \bottomrule
      \end{tabular}
\end{table}

\begin{table}[ht!]
   \centering
    \caption{Aggregation of basic topological characteristics of real-world networks after grouping them according to their meaningful correlation proportion from table \ref{table:new2}, meaningful Jaccard similarity from table \ref{table:rankingSigSimJaccard} and meaningful RBO similarity from table \ref{table:new4}. The basic topological characteristics are $\nu$ is the density, $\zeta$ is the transitivity, and $k_{nn}(k)$ is the assortativity. Two states can be given to the density and other characteristics within a group, either high denoted as high denoted as H or low denoted as L. Two states can be given for assortativity, either positive denoted as P or negative denoted as N. If fluctuations occur within a group, they are denoted as X.}
    \label{table:birdseyeviewofgroups}
      \begin{tabular}{p{7cm}p{0.5cm}p{0.5cm}p{0.5cm}}
      \toprule
    & $\nu$ & $\zeta$  & $k_{nn}(k)$  \\
 
\midrule
\textbf{Correlation Rankings}\\
\midrule
Group 1: Adjective Noun $\rightarrow$ Mammals & H & H & N
\\
Group 2: Physicians $\rightarrow$ PGP  & L & H & P
\\
Group 3: Retweets Copenhagen $\rightarrow$ CS Ph.D. & L & L & N
\\
\midrule
\textbf{Jaccard Rankings}\\
\midrule
Group 1: U.S. Airports $\rightarrow$ World Metal Trade & H & H & N
\\
Group 2: Physicians $\rightarrow$ AstroPh & L & X & X
\\
\midrule
\textbf{RBO Rankings}\\
\midrule
Group 1: Adjective Noun $\rightarrow$ Mammals & X & H & X
\\
Group 2: Les Mis\'erables $\rightarrow$ EuroRoads & L & X & X
\\

    \bottomrule
       
      \end{tabular}

\end{table}

\subsubsection{Clustering the networks based on the evaluation measures sample sets using the $K$-means algorithm}
For a given evaluation measure $\gamma$, a network $S_i$ is represented by a multidimensional vector $\Gamma_i= \{\gamma(\alpha_1,\beta_1), \gamma(\alpha_1,\beta_2),..., \gamma(\alpha_i,\beta_j)\}$ made of the various combinations between the centrality and hierarchy measures. Therefore, these  features can be used in order to categorize the networks using the $k$-means clustering algorithm. The main advantage as compared to the previous experiment is that no threshold is used to cluster the networks. Furthermore, comparisons can be performed with the results of the grouping based on the binarized evaluation measure. The main goal of these experiments is to check if the grouping made previously are consistent. This is why the value of  the number of clusters $k$ in the $k$-means algorithm is not based on an optimization criteria, but on the results of the previous experiment. Accordingly, $k$ is set to 3  for the correlation evaluation measures, and $k$=2 for the similarity evaluation measures (Jaccard and RBO).

Table  \ref{table:CorrClustering} reports the content of the three clusters based on the Spearman correlation evaluation measure. It must be compared to the ranked networks based on the binarized evaluation measures reported in table \ref{table:new2}. The first cluster regroups 10 networks. It has a big overlap with the first  group based on the binarized evaluation measures that contains 8 networks. Both groups have 6 common networks (Zachary Karate Club, Madrid Train Bombings, Les Mis\'erables, World Metal Trade and Adjective Noun). The 4 remaining networks belong to the second group based on the binarized evaluation measures. The third cluster is made of 8 networks. Among these networks 7 are common with the third group of table \ref{table:new2}. The common networks are Mouse Visual Cortex, E.coli Transcription, Yeast Protein, EuroRoads, U.S. Power Grids, CS Ph.D., Retweets Copenhagen, and Human Protein. Globally,  the three clusters uncovered by $k$-means have a high overlap with the grouping based on the ranking of the meaningful correlation evaluation measures. These results confirm our previous findings. 

Let's now turn to the comparison of the two clusters uncovered by the $k$-means algorithm with the two clusters of table \ref{table:rankingSigSimJaccard}. Cluster 1 in table \ref{table:SimClustering} contains 7 networks while the corresponding cluster in table \ref{table:rankingSigSimJaccard} contains 9 networks. They have 7 networks in common  (Mammals, Insects, Birds, Zachary Karate Club, Madrid Train Bombing, World Metal Trade, and U.S. Airports). Those networks are characterized by high density and high transitivity. The second cluster in table \ref{table:SimClustering} contains 19 networks while the corresponding group in table \ref{table:rankingSigSimJaccard} contains 16 networks. They have 16 networks in common. This further illustrates the proximity of the two clustering methods.

Finally, lets compare the two $k$-means clusters  reported in table \ref{table:SimClusteringRBO} to the two groups of table \ref{table:new4}.
Those clusters are based on the $RBO_{p=0.9}$ of top-k nodes evaluation measures. There is a complete overlap between the clusters except for two networks which are AstroPh and Retweets Copenhagen.
To summarize, clusters based on the $k$-means algorithm are very similar to the various groups identified by looking at the relations between the topological properties of the networks and the ranking based on the proportion of meaningful evaluation measures. These results strengthen our understanding of the relationship between the topology of networks and the interaction of centrality and hierarchy measures.

\begin{table}[t!]
   \centering
    \caption{Clusters of the networks $S_i, S_j$ using $k$-means according to their Spearman correlation ($\rho_s$) value across all hierarchy and centrality combinations.}
    \label{table:CorrClustering}    
      \begin{tabular}{p{1.5cm}p{9cm}}
      \toprule
   \textbf{Cluster} & \textbf{Network}  \\
    \midrule
    \textbf{Cluster 1}  & Mammals, Insects, Zachary Karate Club,
      Madrid Train Bombings, Physicians, Adolescent Health,  Les Mis\'erables, World Metal Trade, Adjective Noun, U.S. States  
    \\
    \hline
  \textbf{Cluster 2}  & Birds, Reptiles, Internet Autonomous Systems, U.S. Airports, NetSci, GrQc, AstroPh, Facebook Ego, Facebook Politician Pages, PGP 
    \\
    \hline
   \textbf{Cluster 3}  & Mouse Visual Cortex, E.coli Transcription, Yeast Protein, Human Protein, EuroRoads, U.S. Power Grids, CS Ph.D., Retweets Copenhagen 
    \\

    \bottomrule
       
      \end{tabular}

\end{table}

\begin{table}[t!]
   \centering
    \caption{Clusters of the networks using $k$-means according to their Jaccard Similarity ($J$) value across all hierarchy and centrality combinations.}
    \label{table:SimClustering}
      \begin{tabular}{p{1.5cm}p{9cm}}
      \toprule
   \textbf{Cluster} & \textbf{Network}  \\
    \midrule
    \textbf{Cluster 1}  &  Mammals, Insects, Birds, Zachary Karate Club, Madrid Train Bombings, Physicians, World Metal Trade, Adjective Noun, U.S. Airports
    \\
    \hline
   \textbf{Cluster 2}  & Reptiles, Mouse Visual Cortex, E.Coli Transcription, Yeast Protein, Human Protein, Adolescent Health, Les Miserables, Internet Autonomous Systems, U.S. States, EuroRoads, U.S. Power Grids, NetSci, CS Ph.D., GrQc, AstroPh, Retweets Copenhagen, Facebook Ego, Facebook Politician Pages, PGP
    \\

    \bottomrule
       
      \end{tabular}

\end{table}

\begin{table}[t!]
   \centering
    \caption{Clusters of the networks using $k$-means according to their RBO top-k Similarity ($RBO_{p=0.9}$) value across all hierarchy and centrality combinations.}
    \label{table:SimClusteringRBO}
      \begin{tabular}{p{1.5cm}p{9cm}}
      \toprule
   \textbf{Cluster} & \textbf{Network}  \\
    \midrule
    \textbf{Cluster 1}  & Mammals, Insects, Birds, Zachary Karate, Madrid Train Bombings, Physicians, Les Mis\'erables, World Metal Trade, Adjective Noun, Internet Autonomous Systems, U.S. States, U.S. Airports, U.S. Power Grids, Retweets Copenhagen
    \\
    \hline
   \textbf{Cluster 2}  & Reptiles, Mouse Visual Cortex, E.Coli Transcription, Yeast Protein, Human Protein, Adolescent Health, EuroRoads, NetSci, CS Ph.D., GrQc, AstroPh, Facebook Ego, Facebook Politician Pages, PGP
    \\
    \bottomrule
       
      \end{tabular}

\end{table}

\subsection{Comparing the combinations of hierarchy and centrality measures using the Schulze voting method}
In this section, the Schulze voting method is used in order to rank the output of the evaluations-measures for the various combinations of hierarchy and centrality measures. For a given evaluation measure $\gamma$, each of the 28 networks  under test is considered as a voter $S_i$, and the evaluation measure $\gamma(\alpha_i,\beta_j)$ of hierarchy $ \alpha_i$ and centrality measures $ \beta_j$ are the candidates. Voters rank the candidates in the  descending order of the value of each of the 24 evaluation measures $\Gamma_i= \{\gamma(\alpha_1,\beta_1), \gamma(\alpha_1,\beta_2),..., \gamma(\alpha_i,\beta_j)\}$. As voters rank the candidates in the order from the one that they most want to win to the one they least want to win, the outcome of the voting process is the set of evaluation measures ranked from the most correlated/similar to the least correlated/similar according to the population of voters. This experiment is performed using the three correlation measures (Pearson, Spearman, Kendall Tau) and the 5 similarity measures (Jaccard, RBO top-k at $p$=0.5 and $p$=0.9 and RBO of entire set of nodes at $p$=0.5 and $p$=0.9).

Table \ref{table:SchulzeRankingsCorrelation} reports the ranking based on the 3 correlation measures. The Kendall Tau correlation between the ranking samples is provided in table \ref{table:corrofschulzerankingsKendall}. It can be seen that Pearson is the least correlated with values around 0.6 while Spearman and Kendall Tau rankings are well correlated with a value of 0.8. As the correlation between the rankings of the 3 correlation measures is not so low, a final rank is provided based on their average ranking.
If we refer to this final rank, the most correlated combinations of centrality and hierarchy are LRC $\alpha_l$ and closeness centrality $\beta_c$. It is followed by LRC $\alpha_l$ and local centrality $\beta_l$. The third combination is triangle participation $\alpha_{tp}$ and degree centrality $\beta_d$. Those combinations have been voted as the most correlated across the 28 real-world networks. For the least correlated combinations, the lowest ranks happen to involve betweenness centrality $ \beta_b$ with the nested hierarchies $k$-truss $\alpha_t$ and $k$-core $\alpha_c$ and then comes PageRank  $\beta_p$ with $k$-truss $\alpha_t$.

Table \ref{table:SchulzeRankingsSimilarity} reports the Schulze rankings according to the similarity evaluation measures. Kendall Tau correlation between these rankings are provided in table \ref{table:corrofschulzerankingsKendall}. It can be seen that on average the Kendall Tau correlation of the similarity evaluation measures are more correlated than the correlation evaluation measure. Therefore, the rankings based on the five different similarity evaluation measures are averaged in order to obtain a final ranking in table \ref{table:SchulzeRankingsSimilarity}. According to this ranking, the top 3 combinations are  LRC $\alpha_l$ and closeness centrality $\beta_c$. Then comes the combination between triangle participation $\alpha_l$ and Katz centrality $\beta_k$ followed by LRC $\alpha_l$ and degree centrality $\beta_d$. The least similar combinations also involve betweenness centrality $\beta_b$ with $k$-core $\alpha_c$ and $k$-truss $\alpha_t$, and then comes PageRank $\beta_p$ with $k$-core $\alpha_c$.

Moreover, in  Schulze similarity rankings, it can be seen in table \ref{table:SchulzeRankingsSimilarity} that general flow hierarchy and mixed hierarchy combinations are ranked as top combinations. From rank 1 till rank 11, all combinations involve either LRC flow hierarchy $(\alpha_l)$ or mixed hierarchy triangle participation $(\alpha_{tp})$. On the other hand, low ranked combinations involve nested hierarchies $k$-core $(\alpha_c)$ and $k$-truss $(\alpha_t)$. This is not the case with correlation where ranks are mixed throughout all different hierarchy types. Nevertheless, comparing the Schulze rankings between both correlation and similarity, there are important common characteristics. Both share the same top combination between LRC flow hierarchy and current-flow closeness centrality. Both don't include any nested hierarchy combination in top 3. Finally, both share common measures with low ranking ($k$-core, $k$-truss and betweenness centrality). In addition, neither $k$-core nor $k$-truss appears in the top 10 combinations in similarity rankings.

In summary, according to the vast majority of rankings $k$-core and $k$-truss are the hierarchy measures which are the less similar/correlated with the centrality measures while LRC and triangle participation appears more frequently as the most similar/correlated with centrality measures. Among the centralities, betweenness and PageRank happen to be the least ones occurring in top rankings as well.

\begin{table}[t!]
   \centering
    \caption{Kendall Tau correlation between the rankings of hierarchy and centrality combinations according to the Schulze Method for the the various evaluation measures. Correlation evaluation measures are Spearman ($\rho_s$), Pearson ($\rho_p$), and Kendall Tau ($\tau_b$). Similarity evaluation measures are Jaccard ($J$), RBO top-k at $p$=0.5 and $p$=0.9 ($R_{0.5}^{top}$ and $R_{0.9}^{top}$), RBO of entire set of nodes at $p$=0.5 and $p$=0.9 ($R_{0.5}^{all}$ and $R_{0.9}^{all}$). The 28 real-world Networks are considered as voters.}
    \label{table:corrofschulzerankingsKendall}   
      \begin{tabular}{ccccccccc}
      \toprule

      & $\rho_s$  &  ${\rho_p}$  & $\tau_b$  & $J$ & $R_{0.5}^{top}$  & $R_{0.9}^{top}$ & $R_{0.5}^{all}$ & $R_{0.9}^{all}$\\
    \midrule
    $\rho_s$ & 1 
    \\
    \addlinespace
    $\rho_p$ & 0.63  & 1
     \\
    \addlinespace
     $\tau_b$ & 0.88 & 0.60 & 1 
     \\
     \addlinespace
    $J$ & 0.40 & 0.55 & 0.36 & 1
     \\
     \addlinespace
    $R_{0.5}^{top}$ & 0.25 & 0.42 & 0.19 & 0.81 & 1 
     \\
     \addlinespace
    $R_{0.9}^{top}$ & 0.25 & 0.41 & 0.19 & 0.80 & 0.92 & 1
     \\
     \addlinespace
    $R_{0.5}^{all}$ & 0.35 & 0.52 & 0.31 & 0.76 & 0.78 & 0.80 & 1 
    \\
    \addlinespace
    $R_{0.9}^{all}$ & 0.34 & 0.51 & 0.28 & 0.74 & 0.76 & 0.79 & 0.91 & 1
    \\
 
     \bottomrule
      \end{tabular}
\end{table}

\begin{table}[t!]
   \centering
    \caption{Ranking of the Schulze Method for all possible combinations between hierarchy $\alpha$ and centrality measures $\beta$. The ranks are sorted based on the average ($SC_c$) of all correlation evaluation measures $\gamma$ used: Spearman ($\rho_s$), Pearson ($\rho_p$), and Kendall Tau ($\tau_b$). The 28 real-world networks are considered as voters. The hierarchy measures are $\alpha_{c}$ = $k$-core, $\alpha_{t}$ = $k$-truss, $\alpha_{l}$ = LRC, and $\alpha_{tp}$ = triangle participation. The centrality measures are $\beta_d$ = Degree, $\beta_l$ = Local, $\beta_b$ = Betweenness, $\beta_c$ = Current-flow Closeness, $\beta_k$ = Katz, and $\beta_p$ = PageRank. Note that ties in the output can exist with the Schulze method. In case of ties in the sum of rankings, ranks are broken randomly.}
    \label{table:SchulzeRankingsCorrelation}   
      \begin{tabular}{p{2cm}cccc}
      \toprule

     &  $\rho_s$ & $\rho_p$ & $\tau_b$  & $SC_c$ \\
   \midrule
   
    1. $\gamma(\alpha_{l}, \beta_c)$
    & 
    1
    &
    1
    &
    2
& 4$/$3
     \\
   2. $\gamma(\alpha_{l}, \beta_l)$ 
    & 
    2
    &
    4
    &
    4
   & 10$/$3
     \\
   3. $\gamma(\alpha_{tp}, \beta_d)$
     &
    6
    &
    3
    &
    3
    & 12$/$3
    \\
    4. $\gamma(\alpha_{c}, \beta_d)$
    & 
    3  & 9 & 1 & 13$/$3
     \\
    5. $\gamma(\alpha_{c}, \beta_c)$
    & 
    5  & 5 & 5 & 15$/$3
     \\
    6. $\gamma(\alpha_{l}, \beta_k)$
    & 
    4
    &
    6
    &
    7
& 17$/$3
     \\
    7. $\gamma(\alpha_{tp}, \beta_k)$ 
    & 
    8
     &
    2
    &
    8
& 18$/$3
     \\
    8. $\gamma(\alpha_{c}, \beta_k)$ 
    & 
    7 &  8 & 6 & 21$/$3
     \\
   9. $\gamma(\alpha_{tp}, \beta_b)$ 
    & 
    10
    &
    9
    &
    11
& 30$/$3
     \\
  10. $\gamma(\alpha_{c}, \beta_l)$
    &  10 & 10 & 10 & 30$/$3
     \\
   11. $\gamma(\alpha_{tp}, \beta_l)$
    &
    12
    &
    7
    &
    12
& 31$/$3
     \\
    12. $\gamma(\alpha_{tp}, \beta_p)$ 
    & 
    13
     &
    9
    &
    10
& 32$/$3 
     \\
     
    13. $\gamma(\alpha_{t}, \beta_d)$ 
    & 
    11 &  13 & 8 & 32$/$3
    \\
    14. $\gamma(\alpha_{l}, \beta_d)$ 
    & 
    15
    &
    8
    &
    11
    & 34$/$3
     \\
 
   15. $\gamma(\alpha_{t}, \beta_k)$
    & 
    16
    &
    12
    &
    10
    & 38$/$3
     \\

    16. $\gamma(\alpha_{c}, \beta_p)$
    & 
    14 & 15 & 10 & 39$/$3
     \\
   17. $\gamma(\alpha_{t}, \beta_l)$
    &
    18 & 11 & 13 & 42$/$3
     \\
   18.  $\gamma(\alpha_{t}, \beta_c)$
    &   
    17 &  14  & 13 & 44$/$3
     \\
    19. $\gamma(\alpha_{tp}, \beta_c)$ 
    & 
    19
     &
    17
    &
    9
& 45$/$3
     \\
    20. $\gamma(\alpha_{t}, \beta_p)$
    & 
    19
    &
    18
    &
    14
    & 51$/$3
     \\

 21. $\gamma(\alpha_{l}, \beta_b)$ 
    & 
    20
    &
    16
    &
    15
& 51$/$3
     \\

22. $\gamma(\alpha_{l}, \beta_p)$
    & 
    22
    &
    13
    &
    18
& 53$/$3
    \\
 
   23. $\gamma(\alpha_{c}, \beta_b)$ 
    &  21  & 20   & 16 & 57$/$3
     \\
    24. $\gamma(\alpha_{t}, \beta_b)$ 
    & 
    24 & 21 & 19 & 64$/$3
     \\

     \bottomrule
      \end{tabular}
\end{table}

\begin{table}[t!]
   \centering
    \caption{Ranking of the Schulze method For all possible combinations between hierarchy $\alpha$ and Centrality measures $\beta$. The ranks are sorted based on the average ($SC_s$) of all correlation evaluation measures $\gamma$ used: Jaccard ($J$), RBO top-k at $p$=0.5 and $p$=0.9 ($R_{0.5}^{top}$. $R_{0.9}^{top}$), RBO of entire set of nodes at $p$=0.5 and $p$=0.9 ($R_{0.5}^{all}$ and $R_{0.9}^{all}$). The 28 real-world networks are considered as voters. The hierarchy measures are $\alpha_{c}$ = $k$-core, $\alpha_{t}$ = $k$-truss, $\alpha_{l}$ = LRC, and $\alpha_{tp}$ = triangle participation. The centrality measures are $\beta_d$ = Degree, $\beta_l$ = Local, $\beta_b$ = Betweenness, $\beta_c$ = Current-flow Closeness, $\beta_k$ = Katz, and $\beta_p$ = PageRank. Note that ties in the output can exist with the Schulze method. In case of ties in the sum of rankings, ranks are broken randomly.}
    \label{table:SchulzeRankingsSimilarity}   
      \begin{tabular}{p{2cm}cccccc}
      \toprule
     &   $J$ & $R_{0.5}^{top}$ & $R_{0.9}^{top}$ & $R_{0.5}^{all}$ & $R_{0.9}^{all}$ & $SC_s$ \\
    \midrule
   1. $\gamma(\alpha_{l}, \beta_c)$
    & 
    1 & 1 & 1 & 1 & 1 & 5$/$5
     \\
    2. $\gamma(\alpha_{tp}, \beta_k)$ 
    & 
    4 & 3 & 3 & 2 & 2 & 14$/$5
     \\
     3. $\gamma(\alpha_{l}, \beta_d)$ 
    & 
    5 & 2 & 2 & 2 & 7 & 18$/$5
     \\
   4. $\gamma(\alpha_{l}, \beta_l)$ 
    & 
    2 & 4 & 5 & 4 & 5 & 20$/$5
     \\
   5. $\gamma(\alpha_{tp}, \beta_l)$
    & 
    6 & 6 & 5 & 3 & 4 & 24$/$5
     \\
    6. $\gamma(\alpha_{l}, \beta_b)$ 
    & 
    9 & 5 & 4 & 4 & 3 & 25$/$5
     \\
   7. $\gamma(\alpha_{tp}, \beta_d)$
    & 
    3 & 9 & 5 & 4 & 5 & 26$/$5
     \\
  8. $\gamma(\alpha_{l}, \beta_k)$
    & 
    6 & 7 & 6 & 5 & 6 & 30$/$5
     \\
  9. $\gamma(\alpha_{l}, \beta_p)$
    & 
    8 & 8 & 6 & 6 & 8 & 36$/$5
     \\
  10. $\gamma(\alpha_{tp}, \beta_c)$ 
    & 
    7 & 10 &  7 & 7 & 9 & 40$/$5
     \\
    11. $\gamma(\alpha_{tp}, \beta_p)$ 
    & 
    11 & 11 & 8 & 8 & 10 & 48$/$5
     \\
    12. $\gamma(\alpha_{t}, \beta_d)$ 
    & 
    12 & 13 & 11 & 9 & 13 & 58$/$5
     \\
    13. $\gamma(\alpha_{tp}, \beta_b)$ 
     & 
    16 & 15 & 9 & 10 & 11 & 61$/$5
     \\
    14. $\gamma(\alpha_{t}, \beta_k)$
    & 
    10 & 12 & 10 & 14 & 17 & 63$/$5
     \\
    15. $\gamma(\alpha_{c}, \beta_d)$
    & 
    15 & 17 & 12 & 9 & 12 & 65$/$5
     \\
   16. $\gamma(\alpha_{t}, \beta_l)$
    & 
    13 & 14 & 12 & 13 & 14 & 66$/$5
     \\
    17. $\gamma(\alpha_{c}, \beta_k)$ 
    & 
    14 & 16 & 13 & 12 & 16 & 71$/$5
     \\
   18. $\gamma(\alpha_{c}, \beta_l)$
    & 
    16 & 19 & 17 & 11 & 15 & 78$/$5
     \\
    19. $\gamma(\alpha_{t}, \beta_c)$
    & 
    16 & 15 & 12 & 17 & 20 & 80$/$5
     \\
   20. $\gamma(\alpha_{c}, \beta_c)$
    & 
    15 & 20 & 15 & 15 & 18 & 83$/$5
     \\
    21. $\gamma(\alpha_{t}, \beta_p)$
    & 
    17 & 18 & 14 & 19 & 21 & 89$/$5
     \\
   22. $\gamma(\alpha_{c}, \beta_p)$ 
     & 
    18 & 22 & 19 & 16 & 19 & 94$/$5
     \\
  23. $\gamma(\alpha_{t}, \beta_b)$ 
    & 
    19 & 21 & 16 & 20 & 23 & 99$/$5
     \\
   24.  $\gamma(\alpha_{c}, \beta_b)$ 
    & 
    20 & 23 & 18 & 18 & 22 & 101$/$5
     \\
     \bottomrule
      \end{tabular}
\end{table}

\section{Discussion}\label{sec:discussion}
 
The main goal of  this study is to disentangle the relationship between hierarchy and centrality measures in complex networks. To do so, multiple evaluation measures have been used in order to evaluate the interplay between various prominent centrality measures and hierarchy measures in a representative sample of real-world networks originating from different fields.
In the first set of experiments, for each network $S_i$, an evaluation measure $\gamma(\alpha_i,\beta_j)$ is calculated for each hierarchy  $\alpha_i$ and centrality $\beta_j$ measure combinations. The evaluation measures include 3 correlation measures ($\rho_p,\rho_s,\tau_b$)  and 2 similarity measures ($J, RBO$). Heatmaps representing the values of the evaluation measures for all the combinations associated to each network and to each evaluation measures showed a wide variability between the hierarchy-centrality combinations on one hand and between the networks on the other hand. Nevertheless, typical behaviors have emerged. First of all, the heatmaps of the correlation and the Jaccard evaluation measures are more patchy as compared to RBO. Additionally, a number of network categories appears where the interactions between centrality and hierarchy measures are quite comparable. One can also notice already that betweenness centrality and  $k$-core with $k$-truss hierarchy measures stand out from their competitors. At this stage, it appears that there is an underlying organization yet the inner workings between hierarchy, centrality, and network topology need further investigations.

This leads to the second set of experiments in order to further understand the interplay between hierarchy, centrality, and network topology. The goal of these experiments is to cluster the networks according to a given evaluation measure and to investigate if some topological properties are shared by networks belonging to the same cluster.
First of all, each network is represented by the sample set of a given evaluation measure and the Pearson correlation between the sample sets of the networks two-by-two are computed. Hence, if two networks exhibit a similar behavior, the Pearson correlation between their sample sets is high. Results show that some networks share similar behavior. This appears more frequently using the RBO evaluation measure compared to correlation and Jaccard. Indeed, the heatmap representing the Pearson correlation of the networks is less patchy in this case. At this step, more processing is needed in order to cluster the networks. Two strategies are used. First of all, a threshold is used on the correlation matrix  in order to distinguish meaningful with non meaningful correlation values for each evaluation measures. Then, the number of meaningful correlation values is used to rank the networks. Once the networks are ranked, patterns are searched in the topological properties allowing to cluster the networks. Results clearly demonstrate that the driving force in order to explain the behavior of the various centrality and hierarchy combinations is the density, followed by the transitivity and to a lesser extent assortativity. More specifically, the proportion of meaningful correlation/similarity between hierarchy and centrality combinations is high when both density and transitivity are high. Second, $k$-means clustering algorithm is used where each network is represented by the sample set of a given evaluation measure. The number of clusters is fixed according to the previous experiment. Overall, it appears that the clusters are quite similar. This consistency strengthens our findings.

The goal of the third set of experiments is to investigate the orthogonality of the various hierarchy and centrality measures under test. Indeed, if centrality and hierarchy measures convey the same information their correlation or similarity is pretty high. The Schulze method allows to rank the various combinations from most correlated/similar to least correlated/similar. Results of these experiments according to the various evaluation measures are quite consistent. LRC and current-flow closeness followed by triangle participation and degree centrality are at the top of the ranking. At the other extreme, $k$-core and $k$-truss appear to be the least correlated/similar with betweenness and PageRank centrality. Therefore, the latter should be favoured in order to take advantage of the joint information provided by centrality and hierarchy measures.

Now let's relate the values of the evaluation measures to the topological properties of the networks. Considering correlation as an evaluation measure three clusters emerge while there are only two when similarity is used. The top clusters are characterised by a high number of meaningful correlation/similarity between the various combinations of hierarchy and centrality measures. Furthermore, the networks exhibit high density and high transitivity. In this case centrality and hierarchy exhibit comparable behavior. In other words high density and transitivity seem to induce that both centrality and hierarchy measures carry the same type of information about the network structure. This is particularly true for LRC and triangle participation that are always well correlated with the centrality measures. In contrast, in the clusters characterized by low density and/or low transitivity $k$-truss and to a lesser extent $k$-core behave quite differently than the centrality measures. Indeed, in this situation very low values of similarity/correlation with the vast majority of centrality measures are commonly observed. The $k$-truss nested hierarchy is based on the existence of triangles in the network. In case a network has low transitivity and/or density, it is the first measure to account for this feature. Consequently, this results in a greater divergence with centrality measures.

To summarize, $k$-core and $k$-truss hierarchy and betweenness and PageRank centrality measures tend to be the least correlated and the least similar as compared to the alternative hierarchies and centrality measures under test. This trend is more pronounced with networks characterized by low density and low transitivity. In networks having high density and high transitivity it is more difficult to observe a different behavior of hierarchy measures as compared to centrality measures.

\section{Conclusion}\label{sec:concl}

The aim of this paper is to identify the relationship of hierarchy and centrality, as both measures are used to quantify the importance of a node. Several experiments have been performed using different correlation and similarity evaluation measures on real-world networks originating from various fields.

The first main conclusion is that there is a strong relationship between the hierarchy measures, the centrality measures, and the topological properties of a network. High density and high transitivity tend to bring closer centrality and hierarchy measures in terms of correlation or similarity. Conversely, low values of density and/or transitivity translates into low correlation/similarity between hierarchy and centrality measures.

The second main conclusion is that the most orthogonal combinations (least similar and least correlated) involve the nested hierarchy measures ($k$-core and $k$-truss) and the betweenness and PageRank centrality measures. On the opposite, the flow hierarchy and mixed hierarchy measures (LRC and triangle participation) exhibit a comparable behavior than degree centrality and current-flow closeness centrality measures as measured by correlation and similarity measures.

These results have multiple implications. First of all, they allow to substitute a centrality measure to a hierarchy measure in case they are highly correlated and similar. In contrast, combining both measures can also be a good option in order to take into consideration the complementary information carried by each measure when they are not correlated. All of this must be done, taking into consideration the density and transitivity of the given network.

The results of this study pave the way for further investigation about the relationship between centrality and hierarchy measures considering complementary information on the network topology such as the community structure. Furthermore, it gives clear indications about which hierarchy and centrality measures should be combined in a multidimensional scheme in order to detect influential nodes.



\begin{thebibliography}{10}

\bibitem{baker2013complexity}
Alan Baker.
\newblock Complexity, networks, and non-uniqueness.
\newblock {\em Foundations of science}, 18(4):687--705, 2013.

\bibitem{survey2}
Kousik Das, Sovan Samanta, and Madhumangal Pal.
\newblock Study on centrality measures in social networks: a survey.
\newblock {\em Social Network Analysis and Mining}, 8(1):13, 2018.

\bibitem{ghanem2018centrality}
Marwan Ghanem, Cl{\'e}mence Magnien, and Fabien Tarissan.
\newblock Centrality metrics in dynamic networks: a comparison study.
\newblock {\em IEEE Transactions on Network Science and Engineering},
  6(4):940--951, 2018.

\bibitem{survey1}
Francisco~Aparecido Rodrigues.
\newblock Network centrality: an introduction.
\newblock In {\em A Mathematical Modeling Approach from Nonlinear Dynamics to
  Complex Systems}, pages 177--196. Springer, 2019.

\bibitem{das2014local}
Sima Das.
\newblock On local and global centrality in large scale networks.
\newblock {\em arXiv preprint arXiv:1405.5512}, 2014.

\bibitem{lu2016vital}
Linyuan L{\"u}, Duanbing Chen, Xiao-Long Ren, Qian-Ming Zhang, Yi-Cheng Zhang,
  and Tao Zhou.
\newblock Vital nodes identification in complex networks.
\newblock {\em Physics Reports}, 650:1--63, 2016.

\bibitem{multidimensional}
Carla Sciarra, Guido Chiarotti, Francesco Laio, and Luca Ridolfi.
\newblock A change of perspective in network centrality.
\newblock {\em Scientific Reports}, 8, 12 2018.

\bibitem{ghalmane2019centrality}
Zakariya Ghalmane, Mohammed El~Hassouni, Chantal Cherifi, and Hocine Cherifi.
\newblock Centrality in modular networks.
\newblock {\em EPJ Data Science}, 8(1):15, 2019.

\bibitem{ghalmane2019centrality1}
Zakariya Ghalmane, Chantal Cherifi, Hocine Cherifi, and Mohammed El~Hassouni.
\newblock Centrality in complex networks with overlapping community structure.
\newblock {\em Scientific reports}, 9(1):1--29, 2019.

\bibitem{cherifi2019community}
Hocine Cherifi, Gergely Palla, Boleslaw~K Szymanski, and Xiaoyan Lu.
\newblock On community structure in complex networks: challenges and
  opportunities.
\newblock {\em Applied Network Science}, 4(1):1--35, 2019.

\bibitem{e1}
Herbert~A. Simon.
\newblock The architecture of complexity.
\newblock {\em Proceedings of the American Philosophical Society},
  106(6):467--482, 1962.

\bibitem{zafeiris2017we}
Anna Zafeiris and Tam{\'a}s Vicsek.
\newblock {\em Why We Live in Hierarchies?: A Quantitative Treatise}.
\newblock Springer, 2017.

\bibitem{clark2013handbook}
Alexander Clark, Chris Fox, and Shalom Lappin.
\newblock {\em The handbook of computational linguistics and natural language
  processing}.
\newblock John Wiley \& Sons, 2013.

\bibitem{e9}
Enys Mones, Lilla Vicsek, and Tamás Vicsek.
\newblock Hierarchy measure for complex networks.
\newblock {\em PloS one}, 7:e33799, 03 2012.

\bibitem{e8}
Jianxi Luo and Christopher Magee.
\newblock Detecting evolving patterns of self‐organizing networks by flow
  hierarchy measurement.
\newblock {\em Complexity}, 16:53--61, 07 2011.

\bibitem{e13}
D{\'a}niel Cz{\'e}gel and Gergely Palla.
\newblock Random walk hierarchy measure: What is more hierarchical, a chain, a
  tree or a star?
\newblock {\em Scientific reports}, 5:17994, 2015.

\bibitem{e14}
Ala Trusina, Sergei Maslov, Petter Minnhagen, and Kim Sneppen.
\newblock Hierarchy measures in complex networks.
\newblock {\em Physical review letters}, 92(17):178702, 2004.

\bibitem{e15}
Hongsuda Tangmunarunkit, Ramesh Govindan, Sugih Jamin, Scott Shenker, and
  Walter Willinger.
\newblock Network topologies, power laws, and hierarchy.
\newblock {\em ACM SIGCOMM Computer Communication Review}, 32(1):76, 2002.

\bibitem{garcia2017ranking}
Javier Garc{\'\i}a-Algarra, Juan~Manuel Pastor, Jos{\'e}~Mar{\'\i}a Iriondo,
  and Javier Galeano.
\newblock Ranking of critical species to preserve the functionality of
  mutualistic networks using the k-core decomposition.
\newblock {\em PeerJ}, 5:e3321, 2017.

\bibitem{wang2012truss}
Jia Wang and James Cheng.
\newblock Truss decomposition in massive networks.
\newblock {\em arXiv preprint arXiv:1205.6693}, 2012.

\bibitem{malliaros2020core}
Fragkiskos~D Malliaros, Christos Giatsidis, Apostolos~N Papadopoulos, and
  Michalis Vazirgiannis.
\newblock The core decomposition of networks: Theory, algorithms and
  applications.
\newblock {\em The VLDB Journal}, 29(1):61--92, 2020.

\bibitem{Newman2010}
M.~E.~J. Newman.
\newblock {\em Networks: an introduction}.
\newblock Oxford University Press, Oxford; New York, 2010.

\bibitem{kitsak2010identification}
Maksim Kitsak, Lazaros~K Gallos, Shlomo Havlin, Fredrik Liljeros, Lev Muchnik,
  H~Eugene Stanley, and Hern{\'a}n~A Makse.
\newblock Identification of influential spreaders in complex networks.
\newblock {\em Nature physics}, 6(11):888--893, 2010.

\bibitem{saxena2018social}
Rakhi Saxena, Sharanjit Kaur, and Vasudha Bhatnagar.
\newblock Social centrality using network hierarchy and community structure.
\newblock {\em Data Mining and Knowledge Discovery}, 32(5):1421--1443, 2018.

\bibitem{liu2012control}
Yang-Yu Liu, Jean-Jacques Slotine, and Albert-L{\'a}szl{\'o} Barab{\'a}si.
\newblock Control centrality and hierarchical structure in complex networks.
\newblock {\em Plos one}, 7(9):e44459, 2012.

\bibitem{li2017hierarchical}
Yong Li, Wenguo Li, Yi~Tan, Fang Liu, Yijia Cao, and Kwang~Y Lee.
\newblock Hierarchical decomposition for betweenness centrality measure of
  complex networks.
\newblock {\em Scientific reports}, 7:46491, 2017.

\bibitem{oldham2019consistency}
Stuart Oldham, Ben Fulcher, Linden Parkes, Aurina Arnatkeviciute, Chao Suo, and
  Alex Fornito.
\newblock Consistency and differences between centrality measures across
  distinct classes of networks.
\newblock {\em PloS one}, 14(7), 2019.

\bibitem{e6}
David Lane.
\newblock Hierarchy, complexity, society.
\newblock In {\em Hierarchy in Natural and Social Sciences}, chapter~4, pages
  81--119. Springer, 2006.

\bibitem{e10}
Timothy~FH Allen and Thomas~B Starr.
\newblock {\em Hierarchy: perspectives for ecological complexity}.
\newblock University of Chicago Press, 2017.

\bibitem{harenberg2014community}
Steve Harenberg, Gonzalo Bello, La~Gjeltema, Stephen Ranshous, Jitendra
  Harlalka, Ramona Seay, Kanchana Padmanabhan, and Nagiza Samatova.
\newblock Community detection in large-scale networks: a survey and empirical
  evaluation.
\newblock {\em Wiley Interdisciplinary Reviews: Computational Statistics},
  6(6):426--439, 2014.

\bibitem{yang2015defining}
Jaewon Yang and Jure Leskovec.
\newblock Defining and evaluating network communities based on ground-truth.
\newblock {\em Knowledge and Information Systems}, 42(1):181--213, 2015.

\bibitem{mothe2017community}
Josiane Mothe, Karen Mkhitaryan, and Mariam Haroutunian.
\newblock Community detection: Comparison of state of the art algorithms.
\newblock In {\em 2017 Computer Science and Information Technologies (CSIT)},
  pages 125--129. IEEE, 2017.

\bibitem{Wagenseller2016CommunityDA}
Paul Wagenseller and Feng Wang.
\newblock Community detection algorithm evaluation using size and hashtags.
\newblock {\em ArXiv}, abs/1612.03362, 2016.

\bibitem{ghalmane2019immunization}
Zakariya Ghalmane, Mohammed El~Hassouni, and Hocine Cherifi.
\newblock Immunization of networks with non-overlapping community structure.
\newblock {\em Social Network Analysis and Mining}, 9(1):45, 2019.

\bibitem{ibnoulouafi2018m}
Ahmed Ibnoulouafi, Mohamed El~Haziti, and Hocine Cherifi.
\newblock M-centrality: identifying key nodes based on global position and
  local degree variation.
\newblock {\em Journal of Statistical Mechanics: Theory and Experiment},
  2018(7):073407, 2018.

\bibitem{gupta2016centrality}
Naveen Gupta, Anurag Singh, and Hocine Cherifi.
\newblock Centrality measures for networks with community structure.
\newblock {\em Physica A: Statistical Mechanics and its Applications},
  452:46--59, 2016.

\bibitem{ghalmane2018betweenness}
Zakariya Ghalmane, Mohammed El~Hassouni, and Hocine Cherifi.
\newblock Betweenness centrality for networks with non-overlapping community
  structure.
\newblock In {\em 2018 IEEE workshop on complexity in engineering (COMPENG)},
  pages 1--5. IEEE, 2018.

\bibitem{protein}
Minoo Ashtiani, Ali Salehzadeh-Yazdi, Zahra Razaghi-Moghadam, Holger Hennig,
  Olaf Wolkenhauer, Mehdi Mirzaie, and Mohieddin Jafari.
\newblock A systematic survey of centrality measures for protein-protein
  interaction networks.
\newblock {\em BMC systems biology}, 12(1):80, 2018.

\bibitem{brandes2005centrality}
Ulrik Brandes and Daniel Fleischer.
\newblock Centrality measures based on current flow.
\newblock In {\em Annual symposium on theoretical aspects of computer science},
  pages 533--544. Springer, 2005.

\bibitem{webber2010similarity}
William Webber, Alistair Moffat, and Justin Zobel.
\newblock A similarity measure for indefinite rankings.
\newblock {\em ACM Transactions on Information Systems (TOIS)}, 28(4):1--38,
  2010.

\bibitem{schulze2011new}
Markus Schulze.
\newblock A new monotonic, clone-independent, reversal symmetric, and
  condorcet-consistent single-winner election method.
\newblock {\em Social Choice and Welfare}, 36(2):267--303, 2011.

\bibitem{nr}
Ryan~A. Rossi and Nesreen~K. Ahmed.
\newblock The network data repository with interactive graph analytics and
  visualization.
\newblock In {\em AAAI}, 2015.
\newblock [Online]. Available: \url{http://networkrepository.com}.

\bibitem{latora2017complex}
Vito Latora, Vincenzo Nicosia, and Giovanni Russo.
\newblock {\em Complex networks: principles, methods and applications}.
\newblock Cambridge University Press, 2017.
\newblock [Online]. Available:
  \url{https://www.complex-networks.net/datasets.html}.

\bibitem{de2018exploratory}
Wouter De~Nooy, Andrej Mrvar, and Vladimir Batagelj.
\newblock {\em Exploratory social network analysis with Pajek: Revised and
  expanded edition for updated software}, volume~46.
\newblock Cambridge University Press, 2018.
\newblock [Online]. Available:
  \url{http://vlado.fmf.uni-lj.si/pub/networks/data/esna/default.htm}.

\bibitem{kunegis2014handbook}
J{\'e}r{\^o}me Kunegis.
\newblock Handbook of network analysis [konect--the koblenz network
  collection].
\newblock {\em arXiv preprint arXiv:1402.5500}, 2014.
\newblock [Online]. Available: \url{http://konect.uni-koblenz.de}.

\end{thebibliography}

\end{document}